\documentclass[twocolumn,tighten,numberedappendix]{aastex631}
\usepackage{amsmath}
\usepackage{ulem}
\usepackage{stmaryrd}
\usepackage{enumitem}
\setcitestyle{notesep={ }}

\usepackage{graphicx}% Include Figure~files
\usepackage{dcolumn}% Align table columns on decimal point
\usepackage{bm}% bold math
\usepackage{hyperref}
\usepackage[T1]{fontenc}
\usepackage{scalerel,tikz}
\usepackage{fancyvrb}

\usepackage[flushleft]{threeparttablex}
\usepackage{cprotect}

\newcommand{\bnabla}{\bm{\nabla}}

\newcommand{\be}{\begin{equation}}
\newcommand{\ee}{\end{equation}}
\newcommand{\ba}{\begin{eqnarray}}
\newcommand{\ea}{\end{eqnarray}}
\newcommand{\tr}{\tilde r}
\usepackage{rotating}
\usepackage{rotfloat}
\usepackage{hyperref}

\begin{document}

\title{Sparking Dynamic, Relativistic Magnetic Reconnection in a Twisted Flux Tube}

\correspondingauthor{\\
Michael P. Grehan: \href{mailto:michael.grehan@mail.utoronto.ca}{michael.grehan@mail.utoronto.ca} \\
$^\dagger$NASA Hubble Fellow 
}

\author[0009-0003-1842-192X]{Michael P. Grehan$^*$} 
\affil{Department of Physics, University of Toronto, 60 St. George Street, Toronto, ON M5S 1A7, Canada}
\affil{Canadian Institute for Theoretical Astrophysics, 60 St. George Street, Toronto, ON M5S 3H8, Canada}

\author[0000-0003-4305-5653]{Christopher Thompson$^*$}
\affil{Canadian Institute for Theoretical Astrophysics, 60 St. George Street, Toronto, ON M5S 3H8, Canada}

\author[0000-0002-7301-3908]{Bart Ripperda}
\affil{Department of Physics, University of Toronto, 60 St. George Street, Toronto, ON M5S 1A7, Canada}
\affil{Canadian Institute for Theoretical Astrophysics, 60 St. George Street, Toronto, ON M5S 3H8, Canada}
\affil{D. A. Dunlap Department of Astronomy, University of Toronto, Toronto, ON M5S 3H4, Canada}
\affil{Perimeter Institute for Theoretical Physics, Waterloo, ON N2L 2Y5, Canada}

\author[0000-0001-9199-7771]{James R. Beattie$^{\dagger}$}
\affil{Canadian Institute for Theoretical Astrophysics, 60 St. George Street, Toronto, ON M5S 3H8, Canada}
\affil{Department of Astrophysical Sciences, Princeton University, Princeton, NJ 08544, USA}
\affil{School of Natural Sciences, Institute for Advanced Study, 1 Einstein Drive, Princeton, NJ 08540, USA}

\date{\today}

\begin{abstract} 
    Magnetic reconnection around neutron stars (NSs) forms small-scale currents while transferring energy to electron-positron pairs.  These dissipating currents cover a wide range of scales, potentially leaving an imprint on time-variable X-ray bursts and pulsed radio emission. We present a reduced, planar model of internal reconnection in a magnetic flux tube with relativistic magnetization, finite resistivity, and strong guide field.  Comparing with realistic NS parameters, this model represents the dynamics of a slender magnetic flux tube that is much thinner than the stellar radius. The seed current profile is smooth and does not spontaneously reconnect;  instead, a discrete and localized current perturbation is introduced, representing e.g. localized energy release from a NS crust or magnetosphere. Three high-resolution relativistic magnetohydrodynamics simulations are described:  one has uniform resistivity and Lundquist number such that reconnection proceeds in the plasmoid-mediated regime;  the other two account for the feedback of charge starvation on the resistivity in current layers, where density is depleted; we provide a more general explanation for this depletion. All simulations produce a few dominant magnetic X-points and a secondary Kelvin-Helmholtz instability that excites high-wavenumber currents. We quantify the reconnection rate, the intermittency of dissipation, and the time profile of current spikes.  We further present three-dimensional simulations confirming a limited growth of the ideal kink mode in the strong guide field regime. Applications are discussed, including to radio pulse microstructure and current damping below the magnetar surface during an X-ray flare, leading to mass loss and X-ray afterglow.
\end{abstract}


\section{Introduction}\label{sec:intro}

A longstanding problem in plasma astrophysics centers on the stability of a twisted magnetic field that is coupled to dilute plasma and tied to the surface of a star. Heating of plasma outside the star is facilitated by small-scale current structure and magnetic reconnection \citep{Parker:1972ApJ...174..499P, Pontin2020}.  The study of small-scale reconnection presented here is motivated by the peculiar behavior of magnetars \citep{1992ApJ...392L...9D}, compact stars supporting $\sim 10^{20}$~Ampere currents that show dramatic variability ranging from sub-millisecond to year-long timescales \citep{2017ARA&A..55..261K}. 

Magnetars are especially sensitive probes of current driven instabilities because the magnetic field is the dominant energy source for their radiative emissions, which range across the full electromagnetic spectrum, from the radio to gamma ray bands.  Some magnetars produce peculiar radio pulsations with extreme intermittency and variability (e.g. \citealt{Dai2019,Lower2021,Huang2023}), while others produce repeating, sub-second X-ray bursts that can show strong intraburst variability \citep{Gogus01,Huppenkothen2015}.

The observations of rapid variability in magnetars combine with the problem of plasma heating at
extreme magnetization to motivate the present study of magnetic reconnection on small
scales. We are interested in the dynamics of currents around the kinetic scale 
$d_e = c/\omega_p$, where $\omega_p$ is the plasma frequency and $c$ is the speed of light.  
Outside a neutron star, the transfer of magnetic energy to particles and radiation
is concentrated in zones where the current becomes charge-starved (e.g. in
strongly sheared Alfv\'en waves or small scale current sheets: \citealt{TB98,Nattila22}).   
Except in the most luminous outbursts, charged particles must rapidly relax to the lowest Landau state and form a nearly one-dimensional gas.

This paper considers the dynamics of a slender flux tube with large magnetization $\sigma = B^2/4\pi w \gg 1$, where $w$ is the plasma enthalpy density and $B$ the magnetic field strength.
While global magnetospheric simulations capture the large-scale structure and dynamics of magnetar magnetospheres (e.g. \citealt{
2012PhRvD..86j4035L,
2012ApJ...754L..12P,
2013ApJ...774...92P,
2020ApJ...900L..21Y,
2021ApJ...908..176Y,
Yuan22,
2023ApJ...947L..34M,
2024ApJ...972..139M,
2025ApJ...981L..17M,
2025ApJ...980..222B, 
2025PhRvL.135z5201B,
2025ApJ...995L..57B,
2026ApJ..1005..158G, 
2026ApJ..1006L..45C}), 
they cannot simultaneously resolve the macroscopic scale of a slender flux tube and the kinetic scales relevant to reconnection. 
We therefore elect to perform local simulations, sacrificing a description of the global magnetospheric geometry in favor of 
an improved separation of scales. 

An understudied aspect of the problem 
is the effect of a compact and localized perturbation on a smooth current structure. The magnetic configuration studied in this paper has initial axial symmetry and carries vanishing net current. Spontaneous tearing is too slow to be observed;  instead the perturbation plays a key role in triggering rapid reconnection of
the non-potential magnetic field.  The evolution is studied in two dimensions (2D) at large magnetization ($\sigma \sim 10^2$) by solving the equations of resistive, relativistic magnetohydrodynamics (MHD).  The presence of a local perturbation is motivated in a magnetar by the possibility of a localized yielding in the crust, driven e.g. by fault-like response to large-scale magnetic stresses \citep{Thompson_2017} or plastic flow driven by Hall waves \citep{TD96,Li2016,Lander2019}.

We adopt both a small, uniform resistivity and a recently developed non-uniform resistivity law that accounts for the effects of charge starvation \citep{2025ApJ...978L..45M, Ripperda2026}.   The physical scale at which charge starvation sets in depends strongly on the radiative state of the magnetar. During a bright X-ray burst with super-Eddington luminosity, the plasma may approach local kinetic equilibrium at a temperature $\sim m_ec^2$ \citep{Feroci01};  in a quiescent state, the persistent output of hard X-rays can be accommodated by a flow of $e^\pm$ pairs with density $\sim 10^2$ times the charge starvation value \citep{B13,TK20}.  In the outburst regime, the kinetic scale may then be as small as $d_e \sim 10^{-11}$ cm;  whereas in quiescence, the plasma frequency is inferred to be $\sim 10^{14}$ Hz in the inner magnetosphere, corresponding to $d_e \sim 10^{-3}$ cm. This motivates our dual approach to modeling the resistivity, where we focus on representing as wide a range of hydromagnetic scales as possible, or, alternatively, on directly resolving dissipation at kinetic scales.

\subsection{Plan of the Paper}

The paper is organized as follows. Small-scale twisting of a stellar magnetic field is described qualitatively in Section~\ref{sec:twist}.  The model, numerical methods, and the description of the employed plasma resistivity model are summarized in Section~\ref{sec:initial_cons}. Results are presented in Section~\ref{sec:results}.   The role played by charge starvation of small-scale currents is investigated in Section~\ref{sec:starve}.  Our results are summarized in  Section~\ref{sec:discussion} along with implications for magnetars and pulsars. The Appendices present a complementary three-dimensional (3D) model, describe the approach to mesh refinement and tests of numerical convergence, and review our approach to handling charge-starvation effects using a non-uniform adaptive resistivity law.   

In this paper, we use the notation $X = X_n\times 10^n$ to denote the value of quantity $X$ in c.g.s. units.

\section{Unstable Forcing of Twist in a Stellar Magnetosphere}\label{sec:twist}

We consider the response of a weakly twisted magnetic field, ${\bm B}(x,y) =  B_z\hat z + {\bm B}_\perp$, to a localized perturbation.
Here, $\hat z$ denotes the direction along the background field, while ${\bm B}_\perp$ is the transverse field associated with the twist. Throughout the paper, the subscripts $\perp$ and $z$ denote the transverse and parallel components of vector quantities, respectively, while $\hat{a}$ denotes the unit vector in the direction of an arbitrary vector $\bm{a}$.
The problem is simplified if the weak non-potential magnetic field ${\bm B}_\perp$ is concentrated in the $xy$-plane in a domain of size $\sim R$.  This current-carrying flux rope may be anchored at one or both ends, and in the latter case supports a net twist $\Delta\phi$ along its length $L_z$.  This system has translational symmetry in the $z$-direction, as does the perturbation.

Magnetic reconnection may be ignited by the perturbation, developing on the relatively short timescale
\begin{align}\label{eq:trec0}
    t_{\rm rec} = {R\over v_{\rm rec}} \sim  {R\over \beta_{\rm rec} v_{\rm A}}{B_z\over B_\perp}.
\end{align}
Here, $v_{\rm A} = c\sqrt{\sigma/(1+\sigma)}$ is the Alfv\'en speed. The effective reconnection speed $v_{\rm rec}$ is reduced by a factor $B_\perp/B$, reflecting the reduction of the in-plane Alfv\'{e}n speed \citep{Werner2017ApJ, 2025arXiv250320013G}. The coefficient $\beta_{\rm rec}$ reaches $\sim 0.1$ in the kinetic regime \citep{2017JPlPh..83e7101C, Sironi_2025}. We focus on the case of relativistic magnetization, with $\sigma \equiv B_z^2/4\pi w = \mathcal{O}(100)$ and $v_{\rm A}$ close to the speed of light.  Then the system initially is in nearly force-free equilibrium,
\begin{align}\label{eq:ff}
    {\bm J}\times{\bm B} = {c\over 4\pi} (\bnabla\times{\bm B})\times{\bm B} = 0,
\end{align}
where ${\bm J}$ is the current density. 

Once reconnection is excited in this way, the dynamics has approximate vertical symmetry and the fluid variables vary slowly along the guide field $B_z$.  Even if the flux tube supports a large twist $\Delta\phi = L_z|\bnabla\times{\bm B}_\perp|/2B_z$ over a length $L_z$, and is formally kink unstable, the ideal and resistive kink modes have negligible growth on the timescale $t_{\rm rec}$ \eqref{eq:trec0} when $|B_\perp| \ll |B_z|$. 
The fastest growing linear kink mode is the ideal $m=1$ mode, with growth rate $\Gamma_{\rm kink} \sim 0.15 (v_{\rm A}/R)(B_\perp/B_z)^3$
\citep{2024ApJ...966..173R}.  We then have
\begin{equation}
    t_{\rm rec} \Gamma_{\rm kink} = \frac{\Gamma_{\rm kink}}{\Gamma_{\rm rec}} \sim 
\frac{0.15}{\beta_{\rm rec}} \left(\frac{B_\perp}{B_z}\right)^2
,
\label{eq:rec_vs_kink_rate}
\end{equation}
and therefore $\Gamma_{\rm rec} \gg \Gamma_{\rm kink}$ when $B_\perp \ll B_z$.
The flux tube is then slender when at least mildly twisted $(\Delta\phi \gtrsim 1)$ with $\Gamma_{\rm rec} \gg \Gamma_{\rm kink}$
\begin{align}
    {R\over L_z} \sim {1\over \Delta\phi} {B_\perp\over B_z} \ll 1.
\end{align}

\subsection{Boundary Forcing of a Line-Tied Magnetic Field}

Consider first the case where the magnetic field is tied at two planar boundaries (e.g., $z = 0,\, L_z$); this configuration is a simplified representation of the part of the neutron star magnetosphere with closed magnetic field lines, with the planar boundaries representing the magnetic footpoints. Many numerical studies of the stability of line-tied fields (the ``Parker problem'') find that smooth displacements of the magnetic footpoints on one boundary generate high-wavenumber structure in the vertical current density $J_z$ \citep{Pontin2020}.  The thickness of these current sheets appears to decrease exponentially with time \citep{vanB1986} and nearly force-free structures with tightly wrapping current sheets can be generated \citep{Pontin_Hornig2015}. Some dynamic experiments have found that the 2D Fourier spectrum $J_z(k_\perp)$, for wave vector $\bm{k}$, steepens as the timescale for twisting lengthens in comparison with the magnetospheric Alfv\'en time $L_z/v_{\rm A}$ (e.g. \citealt{Dmitruk2003,Rappazzo2008}); the dependence of this effect on the boundary motions needs further study.

Our focus in this paper is on a small patch of the  twisted magnetosphere. The influence of line tying on reconnection at scale $R \sim 1/k_\perp$ is measured by comparing the distance $v_{\rm A} t_{\rm rec}$ with the length $L_z$ of the field lines.  We find from Equation~\eqref{eq:trec0},
\begin{align}
    {v_{\rm A} t_{\rm rec}\over L_z} \sim  {R\over \beta_{\rm rec}L_z} {B\over B_\perp} \sim 
    {2\over \beta_{\rm rec}}{B\over L_z\,|\bnabla\times{\bm B}|_z(k_\perp)}.
\end{align}
This quantity drops below unity at some scale as long as $J_z$ diverges at large $k_\perp$.  This corresponds to a magnetic power spectrum $B^2 ({k_\perp}) \propto k_\perp^{-\alpha}$ with index $\alpha < 3$.  At scales smaller than this, the process of triggered reconnection becomes approximately planar (Section~\ref{sec:3d_dyn}).

Motions of the magnetar crust may also directly inject localized currents into the magnetosphere.  Measurements of magnetar X-ray outbursts and their afterglow are consistent with the formation of narrow, fault-like features with width less than a kilometer.   The durations of all but the brightest X-ray bursts cluster around $\sim 0.1$~s \citep{Gogus01}; this is comparable to the time for a shear wave to propagate around the neutron star, and is consistent with the global relaxation of elastic stresses during the burst. At the same time, periods of burst activity are followed by extended phases of X-ray afterglow, with an equivalent blackbody area comprising only $\mathcal{O}(10^{-2})$ of the magnetar surface \citep{2017ARA&A..55..261K}.  These properties together suggest the presence of dissipative structures with a large aspect ratio, and imply the direct injection of narrow current layers into the magnetosphere by crustal displacements.

Small-scale twist can also be directly injected into the magnetosphere via a combination of plastic flow and Hall drift.   The flow rate is even more sensitive to fluctuations in applied stress than it is to temperature \citep{Chugunov2010,Thompson_2017}.  Hall drift generates variations in Maxwell stress on scales comparable to or smaller than the crustal thickness (e.g. \citealt{Li2016,GP2019}), which most directly impact the current power spectrum in the magnetosphere.

\subsection{Reconnection Near a Magnetic Separatrix}

The magnetic separatrix of an isolated neutron star (the flux surface dividing open and closed field lines) can support a persistent current sheet. In both pulsars and magnetars, the magnetic twist varies between open and closed field lines, and in particular can vary over a short distance across the separatrix \citep[e.g.][]{
2014ApJ...785L..33P,
2014ApJ...795L..22C,
2017A&A...607A.134C,
2018ApJ...855...94P,
2019ApJ...876L...6P, 
2022ApJ...939...42H,
Bransgrove2023,
2023ApJ...943..105H}. The stability of this strong magnetic shear is of interest here. In a rotation-powered pulsar, the location of the separatrix is determined by the rotation rate of the star; magnetic field lines corotating with the star cannot move faster than the speed of light \citep{GJ69}.  The outgoing flow of corotation charge along open magnetic field lines is surrounded by a thinner return current layer \citep{Bransgrove2023}.  The situation is more complicated (and potentially more time variable) in a magnetar, where the release of energy is dominated by the decay of the magnetic field.

The detection of bright non-thermal X-ray emission from magnetars, combined with strong variations in spindown rate, points to the presence of sustained electric currents flowing through the corotating magnetosphere \citep{TLK02,BT07,B13,TK20}. Magnetic twist can diffuse or reconnect onto the most extended closed magnetic field lines \citep{T08a,B09}, or instead, in sources of bright radio bursts, be injected close to the poles \citep{Yuan22,T23}.

\begin{figure}
    \centering
    \includegraphics[width=\linewidth]{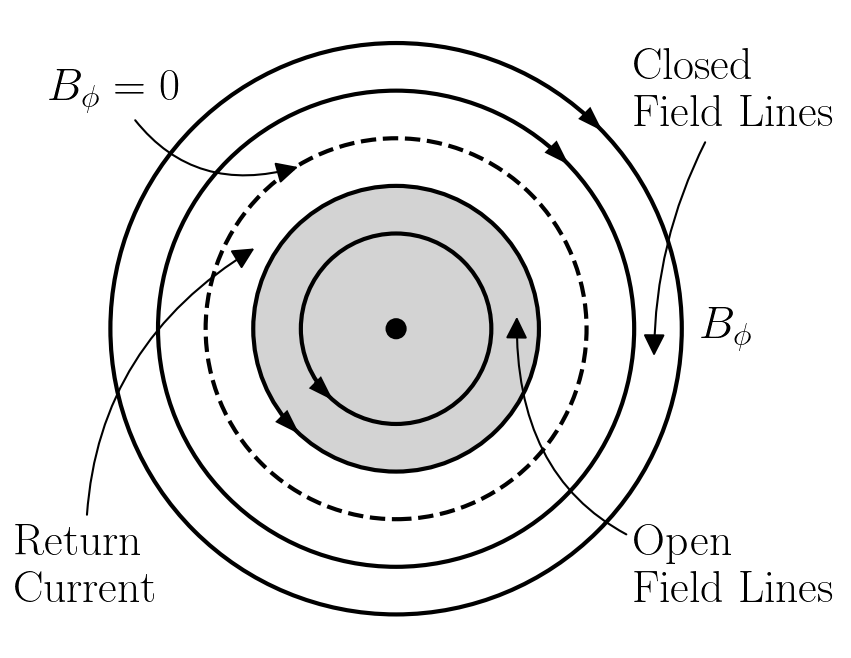}
    \caption{Cross section of magnetic flux surfaces near the dipolar axis of a magnetar
    (adapted from \citealt{TLK02}).
    In one magnetic hemisphere, currents flowing in the closed magnetosphere source a 
    toroidal magnetic field with a sign opposite to the twist on the open flux bundle.
    In radio emitting magnetars, the pulse profile shows strong secular evolution,
    which may be connected with instability of the current near the separatrix.}
    \label{fig:openclosed}
\end{figure}

The magnetospheres of active Galactic magnetars are typically very extended, with spin period $P \sim 1-10$~s and maximum radius up to $cP/2\pi \sim 10^{10}-10^{11}$~cm.  This implies that the magnetic separatrix can contract significantly in response to a relatively weak, persistent flux of magnetohydrodynamic waves or particles near the pole \citep{TB98,Harding99}. In addition, a mismatch must arise between the current flowing along open field lines, where $J/B$ has opposing signs in the two magnetic hemispheres, and the current flowing through the outer magnetosphere, where $J/B$ maintains a uniform sign on each magnetic flux surface (Figure~\ref{fig:openclosed}).  This layer is then a potential site for current-driven instability
\citep{T08a}.

\subsection{Interaction with the Magnetar Surface}

A current sheet formed by small-scale reconnection within the magnetosphere extends a finite distance $v_{\rm A} t_{\rm rec} \sim ct_{\rm rec}$ along the background magnetic field.  This structure may be viewed as a superposition of oppositely moving Alfv\'en waves \citep{2016ApJ...827L..28H, Ripperda2021_FFR}. These waves develop a long tail as the current profile in the reconnection zone settles down to a new equilibrium value.  This process is expected to occur throughout the volume of the magnetosphere, as energy deposited in large-scale magnetic shear is transferred to higher-$k_\perp$ modes and eventually to kinetic scales.  

Two major considerations for this process are: (i) the potential role of the magnetar crust in restricting hydromagnetic displacements of field lines; and (ii) the deposition of magnetic energy below the surface of the magnetar, at depths $\Delta z > ct_{\rm rec} \gg 1/k_\perp$.  We consider the first issue here and defer a discussion of the second to Section~\ref{sec:surface}.

A simple energetic argument suggests that magnetic reconnection within a slender flux tube is not significantly suppressed by line tying of the footpoints. Consider a configuration similar to the one investigated here, a twisted magnetic field $\{B_\phi(r),\,B_z(r)\}$ that is initially axisymmetric and in force-free equilibrium \eqref{eq:ff}. The toroidal field $B_\phi$ may reverse sign at some radius $r_0$; we suppose that the reversal is concentrated in a very thin radial shell, with $B_\phi = +B_0$ at $r > r_0$ and $B_\phi = -B_0$ at $r < r_0$.

The energy that is dissipated by reconnection of the toroidal field ought to be partly compensated by the bending of the vertical field that remains tied at $z = 0,\, L_z$.  The toroidal field energy released within the current sheet is expressed in terms of the net angle through which the vertical field is twisted, $\Delta\phi = (L_z/r_0)(B_0/B_z)$,
\begin{equation}
    \mathcal{E}_\phi = L_z\int_{r_0-\Delta r}^{r_0+\Delta r} 2\pi r dr {B_\phi^2\over 8\pi} =
    \Delta\phi^2B_z^2{\Delta r r_0^3\over 2L_z}.
\end{equation}
The toroidal field carried by flux elements at radius $r_0 \pm \xi_r$ may reconnect only if these elements are displaced in the radial direction by $\xi_r \sim 
\pm f_r (r-r_0)$ at the mid-point of the flux rope, with $f_r \gtrsim 2$. The associated perturbation 
as constrained by line tying at the boundaries is $B_r = (\pi\xi_r/L_z)\cos(\pi z/L_z)B_z$. Here, we have taken the fastest growing mode ($m=1$) \citep{2024ApJ...966..173R}, and made use of Faraday's law along with axial symmetry to write $B_r$.
The energy associated with the perturbation is
\begin{equation}
    \mathcal{E}_r = \int_0^{L_z}\int_{r_0-\Delta r}^{r_0+\Delta r}2\pi r dr  dz{B_r^2\over 8\pi} = f_r^2B_z^2{\pi^2\Delta r^3 r_0\over 12 L_z}.
\end{equation}
The ratio is $\mathcal{E}_r/\mathcal{E}_\phi = (1/6)(\pi f_r/\Delta\phi)^2(\Delta r/r_0)^2$, which is less than unity for $\Delta\phi \gtrsim \pi$ even if the field reversal is smooth, $\Delta r \lesssim r_0$.   In this case, the
sideways displacement of the reconnecting field lines is not impeded by line tying.

\begin{figure*}
    \centering
    \includegraphics[width=1\linewidth]{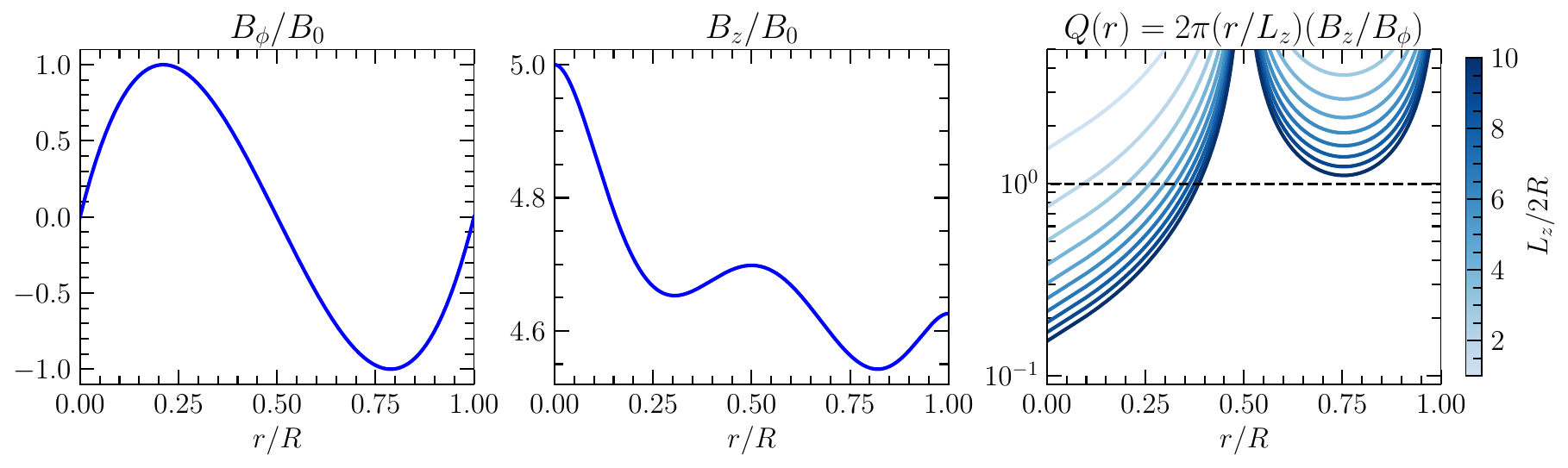}
    \caption{Profiles of the in-plane magnetic field ${\bm B}_\perp = B_\phi(r) \hat \phi$
    (left) and guide field $B_z(r)$ (middle) in axisymmetric force-free
    equilibrium \eqref{eq:ff}, for $B_g = 5 B_0$.  In-plane field and current vanish for $r > R$, where $B_z$ is constant.  Right panel shows the safety factor $Q$ that is implied by
    projecting the 2D field profile into a third dimension with length $L_z$ and aspect ratio 
    $1 \leq L_z/2R \leq 10$. Note that $Q$ diverges as $B_\phi \rightarrow 0$ at $r=R/2$ and $r=R$. 
    A black dashed line indicates the threshold for kink instability ($Q=1$).}
    \label{fig:profiles}
\end{figure*}

\section{Model and Numerical Methods}\label{sec:initial_cons}

\subsection{Magnetic Configuration}

We evolve a relativistic magnetohydrodynamic fluid from a cylindrically symmetric magnetic configuration $\{B_\phi(r),B_z(r)\}$ in 2D, in response to a strong local perturbation. The out-of-plane current $I_z(r) = \int_0^r 2\pi r' dr' J_z(r')$ vanishes outside some radius $r = R$ and we consider the case where the toroidal field $B_\phi$ reverses sign.  The current density remains finite at a small radius, allowing us to write $B_\phi(r) = (r/R)(1-r/R) f(r)$ over $0 < r < R$, for some smooth, non-singular function $f(r)$.  

The default field configuration, shown in Figure~\ref{fig:profiles}, has $B_\phi$ reverse sign once at radius $R/2$,
\begin{equation}
     B_\phi(r) = \left\{
     \begin{array}{cc}
         K_{\phi 1} B_0 \displaystyle\left({r\over R}\right)\left(1-{r\over R}\right)
    \left({1\over 2} - {r\over R} \right), & r\leq R, \\
       0,  & r > R.
    \end{array}
    \right.
    \label{eq:Bphi1}
\end{equation}
The toroidal field peaks at $B_\phi = \pm B_0$ for a value of the coefficient  $K_{\phi 1} = 12\sqrt{3}$.

The guide field $B_z$ is parameterized by its peak value at $r=0$;  we take
\begin{equation}
    B_g \equiv B_z(0) = 5B_0.
\end{equation}
Analytic expressions for $B_z$ and the vector potential $A_z$ are readily obtained and used to define the initial magnetic configuration in the simulations (Section~\ref{sec:sims} and  Appendix~\ref{app:Configurations}). The fluid has the equation of state of a relativistic ideal gas with adiabatic index $\hat\gamma = 4/3$ and central magnetization $\sigma =B_g^2/4\pi w = 100$. 

In simulations with uniform resistivity $\eta$, we adopt a value low enough that the plasmoid instability is triggered, with corresponding Lundquist number \citep{2025arXiv250320013G}
\begin{equation}
    S_\perp = { (\pi R/2) v_{\rm A}\over \eta c^2/4\pi} {B_0\over B_g} \sim  10^5.    
\end{equation} 
This cylindrically symmetric state is perturbed by applying a strongly localized twist, centered at $(r_0 = R/2,\,\phi_0 = 5\pi/4)$, with Gaussian width $\Delta/ 2 R = 0.01$, amplitude $\tilde{A}/2RB_0 = 0.01$, and profile
\begin{equation}\label{eq:Azpert}
    A_z^1 = \tilde{A} \exp \left( - \frac{r^2 + r_0^2 - 2 r r_0 \cos(\phi - \phi_0)}{2 \Delta^2} \right).
\end{equation}
When this 2D configuration is extended uniformly into a third dimension over a distance $L_z$, we may compute an equivalent safety factor $Q(r) = 2\pi B_z r/B_\phi L_z \equiv 2\pi/\Delta\phi$. This is shown in Figure~\ref{fig:profiles} for a range of aspect ratios $L_z/2R$. Zones with $Q  < 1$ support a net winding $\Delta\phi > 2\pi$ and are unstable.

For the strong guide field considered in this paper ($B_z \sim 5 B_\perp$), the bulk of the 3D flux tube maintains $Q>1$ as long as $L_z/2R \lesssim 10$. It should also be noted that, even if the kink instability is present, reconnection triggered by a local perturbation can still develop more rapidly than the fastest-growing kink mode (see Equation~\eqref{eq:rec_vs_kink_rate}).  The relative growth rate of the kink mode depends on the ratio $B_\perp/B_z$.

The main text of the paper focuses on the results of 2D simulations of reconnection.
The behavior of the projected 3D structure has also been tested directly in ideal MHD simulations. In Appendix~\ref{app:3D}, we describe two 3D simulations with vertical dimension $L_z = 10R$, the first with a strong guide field ($B_g = 5 B_0$) matching the value taken in the 2D simulations described in the main text, and the second with a moderate guide field ($B_g = 2.25 B_0$). The 3D evolution with strong guide field closely reproduces the 2D simulations, with the $m=1$ kink mode showing negligible growth. When $B_g$ is reduced to $2.25B_0$, the kink mode now becomes noticeable in the 3D evolution when reconnection is sparked by a local perturbation.  Further details are provided in Section~\ref{sec:3d_dyn} and Appendix~\ref{app:3D}.
\begin{figure*}
    \centering
    \includegraphics[width=\linewidth]{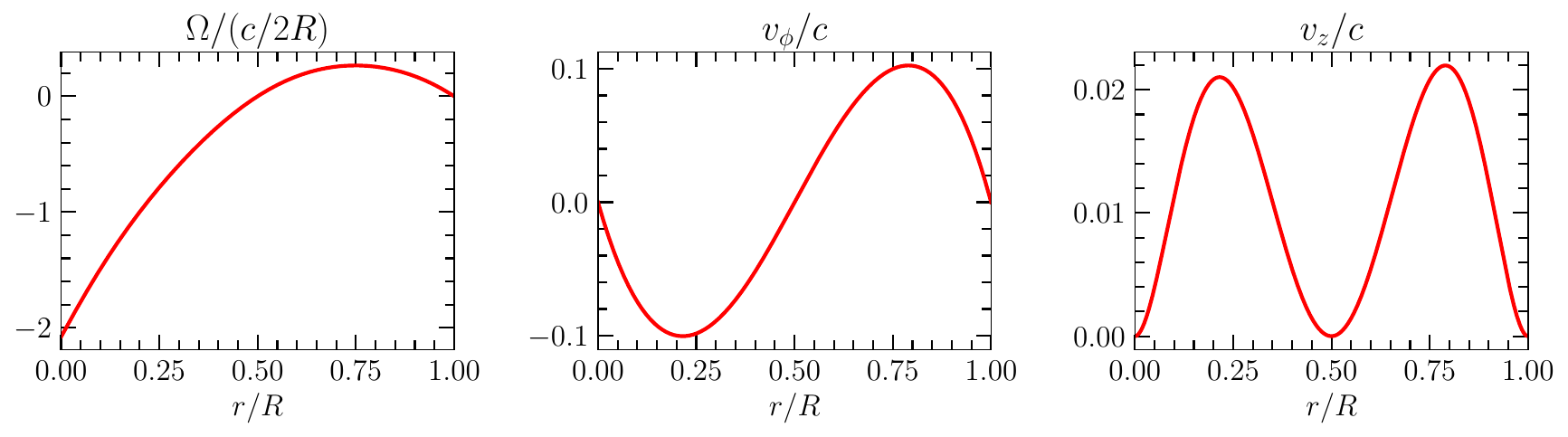}
    \caption{Profiles of the angular velocity $\Omega(r)$
    (left), in-plane velocity $v_\phi(r) = \Omega r$ (middle) and out-of-plane velocity $v_z(r)$ 
    (right) in an axisymmetric force-free equilibrium state with $B_g = 5B_0$ and electric fraction
    $\epsilon = 0.5$ \eqref{eq:FFE_rot}.  The motion vanishes at $r > R$ along with
    the in-plane fields.}
    \label{fig:profiles_rot}
\end{figure*}

\subsection{Background Differential Rotation}

We further consider and simulate the case where the interior of the flux tube is differentially rotating (such an effect is expected when the polar region of a magnetar is actively twisted and field lines near the magnetic separatrix are displaced). The condition of force-free equilibrium in the background is modified by the presence of a radial electric field, $E_r$, representing toroidal ${\bm E}\times{\bm B}$ drift of the magnetofluid,
\begin{equation}\label{eq:FFE_rot}
    \frac{\partial}{\partial r} \left( \frac{B_\phi^2 - E_r^2 + B_z^2}{8\pi} \right) + \frac{B_\phi^2 - E_r^2}{4\pi r} = 0.
\end{equation}
We consider the case $E_r = \epsilon B_\phi$, with $\epsilon$ constant.
The guide field maintains the same profile as in the non-rotating case, with the change in normalization $B_0^2 \rightarrow (1-\epsilon^2) B_0^2$. The background rotation velocity is
\begin{equation}
    \bm{v}/c \simeq - \epsilon \left(\frac{ B_\phi}{B_z} \right) \hat{\phi} + \epsilon \left( \frac{ B_\phi}{B_z}\right)^2\hat{z},
\end{equation}
to leading order in $B_\phi/B_z$.  Figure~\ref{fig:profiles_rot} shows the in-plane and out-of-plane velocity components $\{v_\phi,\,v_z\}$ and the angular velocity $\Omega = v_\phi/r$ for $\epsilon = 0.5$ and $B_g = 5B_0$.

\subsection{Non-Uniform Resistivity}\label{sec:non-uni}

Simulations of resistive MHD with uniform $\eta$ provide a well-controlled environment for studying reconnection dynamics, but they do not capture much of the physics of plasma heating, which occurs on kinetic scales, nor the reconnection rate and current sheet morphology seen in kinetic simulations \citep{Ripperda2026}. At large magnetization, it is important to account for the effects of charge starvation. Free charges are accelerated to relativistic energies as the current density approaches the limiting conduction current. This effect has long been appreciated to play a role in the magnetospheres of pulsars and black holes \citep{GJ69,BZ1977,2001ApJ...547..437L,US14}, and also operates in strongly turbulent and magnetized
flows \citep{TG14,Chen2022,Nattila22}.

The kinetic effect of charge starvation operates on small scales in the high-$\sigma$, pair-plasma regime studied in this work \citep{Goodbred2022_chargestarvation}. To see how the reconnection dynamics are modified by this process, we also carry out simulations that aim to spatially resolve charge-starved current layers. Recently, \cite{Ripperda2026} demonstrated that a non-uniform resistivity model originally derived by \cite{2025ApJ...978L..45M}, which was calibrated to the physics of charge-starved magnetic X-points, yields reconnection rates and macroscopic flow properties comparable to those obtained in fully kinetic particle-in-cell (PIC) simulations in the high-$\sigma$ regime.  This agreement is agnostic to field geometry and strength of the guide field
\citep{Ripperda2026}. 

The resistivity is derived by setting the advective charge flow $q \bm{v}\rightarrow 0$ in the current density (here $q$ is charge density) and by assuming that the current density near the X-point is charge-starved, $|\bm{J}| \sim n ec$, giving
\begin{equation}\label{eq:eta}
    \eta= \frac{|\bm{E}^*|}{n e c}.
\end{equation}
Here, $n$ is the total density of electrons and positrons and $\bm{E}^* = \Gamma[ \bm{E} + \bm{v}\times\bm{B}/c - (\bm{E}\cdot\bm{v})\bm{v}/c^2]$ when the fluid is in motion with velocity ${\bm v}$ and Lorentz factor $\Gamma$ \citep{1993PhRvL..71.3481B, 2003ApJ...589..893L, 2007MNRAS.382..995K}. Rapid plasma heating in charge-starved zones is captured by growing $\eta$ where ${\bm E}^*$ is strong and $n$ is depleted. Magnetic flux freezing is retained in the surrounding plasma, while the thinning of reconnecting current sheets becomes self-limiting through the charge-starvation process.

We consider an equilibrium plasma state approaching the threshold for charge starvation, $en_0c/j_0 \equiv f_{\rm cs} \sim 10$. From Equation~\eqref{eq:Bphi1}, the peak current density (reached at $r = 0$) is $j_0 \sim c K_{\phi 1} B_0/4\pi R$; the charged particle density $n_0 = \rho_0/m$ is obtained from the (nearly uniform) initial particle mass density $\rho_0$, given a particle mass $m$.  The charge starvation length is then
\begin{equation}
    \Delta_0 = \frac{B_0}{4\pi e n_0} \sim \frac{B_0}{4\pi}\frac{c}{f_{\rm cs} j_0} = \frac{ R}{f_{\rm cs} K_{\phi 1}}.
\end{equation}
We set the current sheet thickness by parameterizing the non-uniform resistivity 
\begin{equation}\label{eq:etamod0}
    \frac{\eta}{4\pi} = t_c \left( \frac{\Delta_0}{2R} \right) \left( \frac{|\bm{E}^*|}{B_0}\right) \left( \frac{\rho_0}{\Gamma \rho} \right);\;\quad t_c \equiv{2R\over c},
\end{equation}
as in \citet{Ripperda2026}. Adopting a baseline uniform resistivity $\eta_0$, the non-uniform resistivity is numerically implemented as
\begin{equation}\label{eq:etamod}
     % \eta = \max (\eta_0, \eta^*).
   \frac{\eta}{4\pi} = \max \left\{ \frac{\eta_0}{4\pi}, t_c \left( \frac{\Delta_0}{2R} \right) \left( \frac{|\bm{E}^*|}{B_0}\right) \left( \frac{\rho_0}{\Gamma \rho} \right)\right\},
\end{equation}
with typically $\eta \gg \eta_0$ in reconnecting current sheets.

\subsection{Numerical Simulations}\label{sec:sims}

\subsubsection{Numerical methods, codes and governing equations}

All the (2D) simulations presented in the main text (as summarized in Table~\ref{tab:sims}) are numerical solutions to the equations of resistive, relativistic MHD in flat spacetime. They are performed using the Black Hole Accretion Code (\Verb+BHAC+) \citep{Porth:2016rfi, Olivares2019,Ripperda:2019lsi}.
The \texttt{BHAC} simulations employ the TVDLF approximate Riemann solver, a first-second order Runge-Kutta scheme in which the electric field is solved implicitly \citep{2013MNRAS.428...71B,Ripperda:2019lsi}, and the LIMO3 reconstruction method \citep{2009JCoPh.228.4118C}. The 3D simulations presented in Appendix~\ref{app:3D} assume an infinitely conductive plasma (i.e. ideal MHD). They are performed using \texttt{AthenaK} \citep{2024arXiv240916053S, 2025ApJS..278...50Z, 2025ApJS..276...35F}, a GPU-accelerated code that, for our purposes, solves the ideal, relativistic MHD equations with a finite volume approach. The \texttt{AthenaK} simulations employ the HLLE approximate Riemann solver, and 2nd order Runge–Kutta time integrator coupled with the PPMX reconstruction method \citep{MCCORQUODALE2015181}.  

All times are recorded in units of the flux tube light-crossing time, $t_c = 2R/c$. All simulations are initiated with uniform density $\rho_0$, uniform dimensionless temperature $T_0 = p_0/\rho_0 c^2 = 1$, and magnetization $\sigma = 100$ at $r=0$. The resistivity in uniform-$\eta$ simulations and the baseline resistivity in the simulations with modified resistivity law \eqref{eq:etamod}  are both set to $\eta_0/4\pi = 10^{-6}\,t_c$.

The adoption of ideal MHD in the 3D simulations is motivated purely by computational constraints. In such simulations, the representation of current structures is limited by the grid scale, whereas resolving a current sheet in resistive MHD requires approximately a factor of 10 higher resolution. Thus, achieving the asymptotic (plasmoid-mediated) regime with ideal MHD requires $\sim 10$ times less resolution than a well-resolved resistive simulation. 
We note that ideal MHD simulations have been found to reproduce the macroscopic properties of 2D magnetic reconnection with a uniform resistivity, including the reconnection rate and the generation of tearing and plasmoids at high numerical Lundquist number \citep{2025arXiv250320013G}. Consequently, ideal MHD provides a sufficient description for the 3D simulations, which are designed 
to compare the growth of spontaneous reconnection and the kink instability, as predicted by Equation~\eqref{eq:rec_vs_kink_rate}.

\begin{table}[t]
\centering
\caption{Simulation parameters}\label{tab:sims} 
\begin{ruledtabular}
\begin{tabular}{lccl}
 Simulation & Resolution & Rotation  & Resistivity \\
 ID &  $N_\perp^2$ & $\epsilon = E_\perp/B_\perp$  & Model \\
\hline
\texttt{Rev2D} & $65,\!536^2$ & 0.0 & uniform \\
\texttt{Rev2DNonUni} & $16,\!384^2$ & 0.0  & non-uniform \\
\texttt{Rev2DNonUniRot} & $16,\!384^2$ & 0.5  & non-uniform \\
\end{tabular}
\end{ruledtabular}
\begin{tablenotes}[para]
    \textit{\textbf{Notes.}} Simulation parameters for the initial conditions described in Section~\protect\ref{sec:initial_cons} and Appendix~\protect\ref{app:Configurations}. The mesh refinement employed in these simulations is described in Appendix~\protect\ref{app:amr_res}. All simulations have guide field $B_g/B_0 = 5$, magnetization $\sigma=100$, and domain $|x|,\, |y|\leq 2R$, in 2D.
    \textbf{Column (1):} the unique simulation ID.
    \textbf{Column (2):} the number of cells in each in-plane direction $N_\perp$.
    \textbf{Column (3):} The ratio of the in-plane electric field to in-plane magnetic field, $\epsilon$.          
    \textbf{Column (4):} the resistivity model employed in each simulation.
\end{tablenotes}
\end{table}

\subsubsection{Grid and boundary conditions}

The model flux structure of radius $R$ is centered in a square domain of size $|x|, |y| \leq 2R$.  The in-plane boundary conditions are periodic, but the domain is large enough that the computed dynamics are largely isolated from boundary effects. For 3D simulations, the $z$-boundaries impose a zero gradient condition. This removes line tying at one or both ends of the flux tube, allowing us to model a small patch of a much longer structure that would otherwise be computationally prohibitive. The mesh refinement techniques used in the 2D and 3D simulations are described in Appendix~\ref{app:amr_res}. The 2D simulation $\texttt{Rev2D}$ with uniform $\eta$ is also repeated for a range of resolutions ($8,\!192^2$-$65,\!536^2$ cells), as described in Appendix~\ref{app:con}.  This allows us to test the resolution dependence of the reconnection geometry (the presence of plasmoid instability) and the collective properties of the flow.  Here, the current sheet is resolved with $\sim 10$ 
cells at the maximum resolution ($65,\!536^2$), as presented in the main text.
For the simulations with non-uniform $\eta$, $\Delta_0$ is resolved by $\sim 20$ cells.

\begin{figure*}[h]
    \centering
    \includegraphics[width=0.94\linewidth]{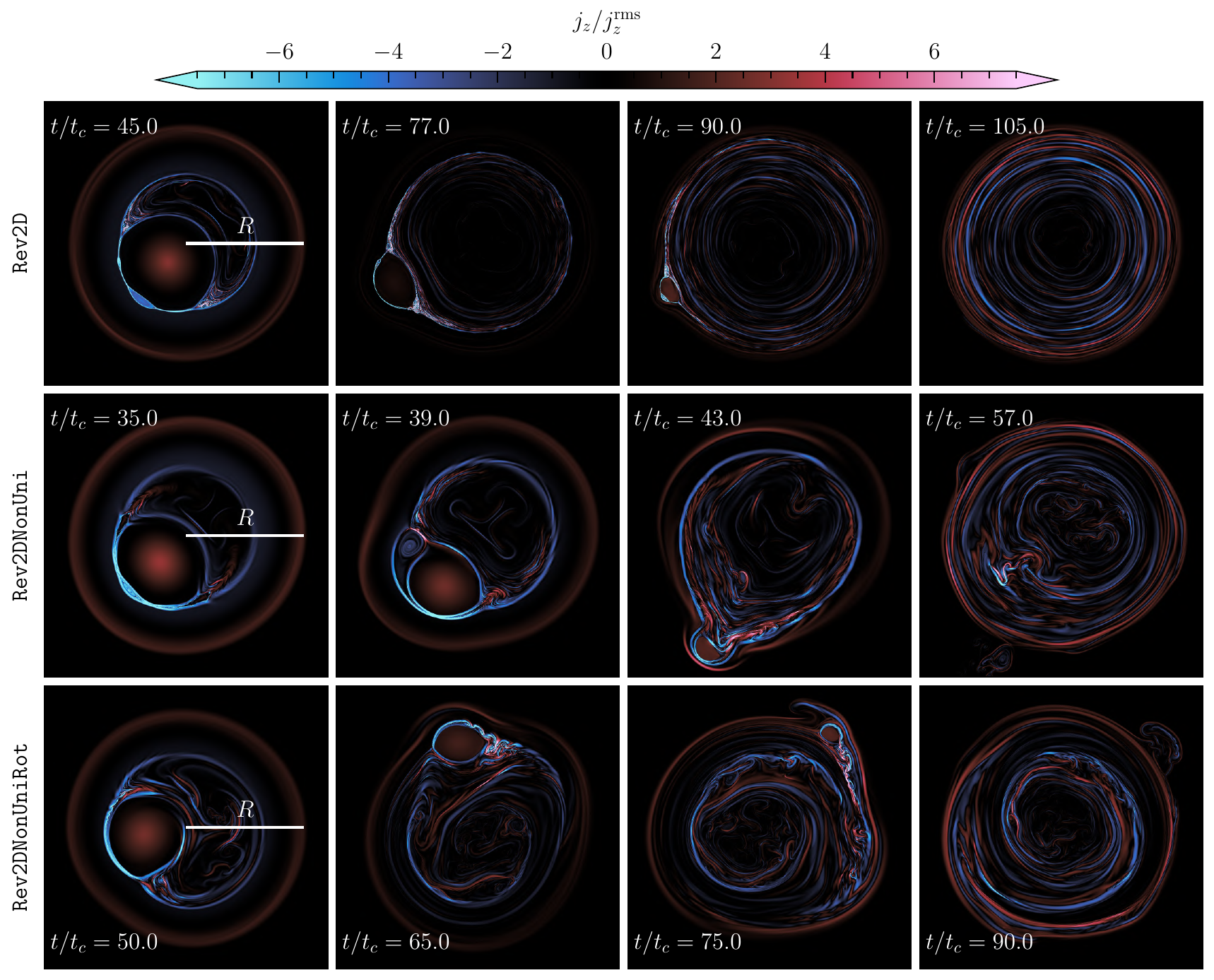}
    \caption{Time evolution of $j_z = (c/4\pi)(\bnabla\times\bm{B})_z$, in simulations \texttt{Rev2D}, \texttt{Rev2DNonUni}, and \texttt{Rev2DNonUniRot}, as normalized by the root-mean-square (r.m.s.) value in the plotted domain $|x|, |y| \leq 6R/5$. In all three simulations, rapid internal reconnection between domains of opposing $B_\phi$ develops at a current sheet in the lower left.  This sheet is unstable to forming plasmoids. Ejecta from the current sheet become turbulent, filling the outer tube with an alternating sequence of weaker current sheets as the inner twist is dissipated. The main current sheet is thicker in simulation \texttt{Rev2DNonUni} and ejects large-amplitude plasmoids at intervals. In the rotating simulation \texttt{Rev2DNonUniRot}, the reconnection process is supplemented by the shear-driven formation of vortices.
    }
    \label{fig:jz_panel}
\end{figure*}

\section{Results}\label{sec:results}

We now describe the evolution of the 2D simulations \texttt{Rev2D}, \texttt{Rev2DNonUni}, and \texttt{Rev2DNonUniRot} (see Table~\ref{tab:sims}). The flux structure consists of two distinct zones with opposing currents: an inner zone inside radius $R/2$ where $I_z > 0$, and an outer zone with a toroidal field of opposing sign where $I_z < 0$ (Figure~\ref{fig:profiles}). The outer and inner currents repel each other, meaning that the inner zone is unstable to displacement of a finite radius. The inner current eventually dissipates, since it supports a lower toroidal flux.

The localized perturbation \eqref{eq:Azpert} that is introduced at radius $R/2$ (where $B_\phi$ switches sign) triggers the formation of a thin, plasmoid-unstable current sheet.  Reconnection is forced by the sideways motion of the inner flux tube (shown in Figure~\ref{fig:jz_panel}), which maintains an approximately circular cross section as it decreases in volume. The end state is a nearly axially symmetric structure with a nested sequence of weaker, circular current sheets (see the right panels of Figure~\ref{fig:jz_panel}).  Differences between the high-$S_\perp$, uniform-$\eta$ simulation \texttt{Rev2D} and the two simulations \texttt{Rev2DNonUni}, \texttt{Rev2DNonUniRot} with modified resistivity, along with the effects of differential rotation, are described in the remainder of this Section.

Note that the triggering of reconnection is sensitive to the initial placement of the perturbation. The result is limited to a dissipating hydromagnetic pulse when the perturbation is displaced from the surface $B_\phi =0$ by more than its width $\Delta$. Indeed, the system appears stable when evolved without perturbation for a duration $\sim 200\,t_c$; a similar result is seen for a geometrically thick 2D Harris sheet with guide magnetic field, which avoids tearing and relaxes  into a self similar current sheet which  Ohmically decays \citep{2007MNRAS.382..995K, 2025arXiv250320013G}.  

\subsection{Global Evolution}

\begin{figure*}
    \centering
    \includegraphics[width=\linewidth]{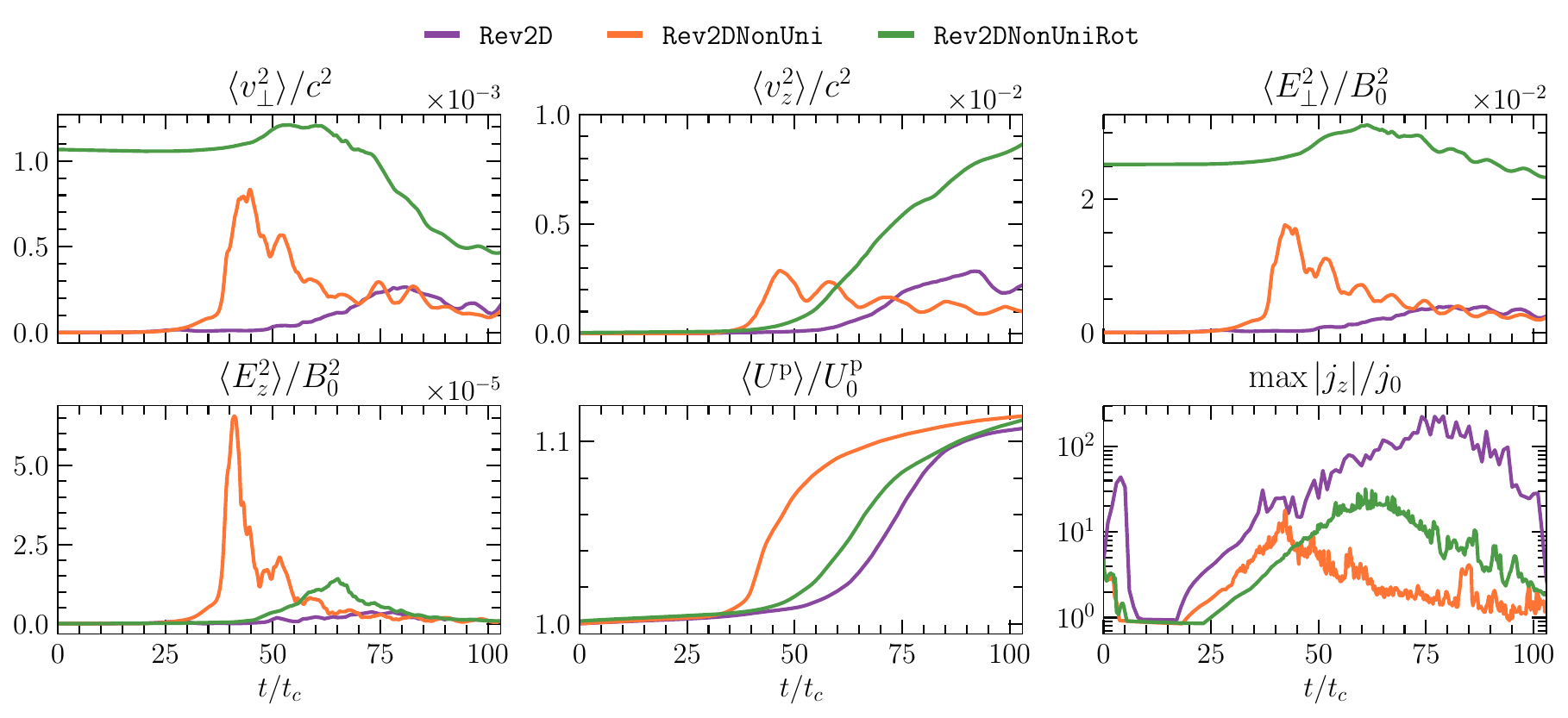}
    \caption{Time evolution of domain-averaged quantities and the maximum current density in simulations \texttt{Rev2D}, \texttt{Rev2DNonUni}, and \texttt{Rev2DNonUniRot}. Reconnection is accompanied by a surge of in-plane motions $v_\perp$ (top left panel) and persistent flow in the out-of-plane direction (top middle panel). In-plane electric energy is related to the in-plane motion via $\langle E_\perp^2\rangle/B_0^2 \simeq (B_g/B_0)^2\langle v_\perp^2/c^2\rangle = 25\langle v_\perp^2/c^2\rangle$ (top right panel). Out-of-plane electric energy is dominated by the non-ideal electric field $\langle E_z^2\rangle/B_0^2 \simeq \eta^2 \langle J_z^2/B_0^2 \rangle$ (bottom left panel). The plasma energy density $U^{p} = \rho h \Gamma^2 c^2 - p$ surges as the in-plane magnetic field exponentially decays and then plateaus (bottom middle panel). The maximum current peaks during reconnection and then rapidly falls off in the late-time evolution (bottom right panel).
    }
    \label{fig:time_avg}
\end{figure*}

The reconnection event is accompanied by a surge in dissipation, which is fastest in model \texttt{Rev2DNonUni}.  Here the reconnection rate rises to $\beta_{\rm rec} \sim 0.1$, driven by the non-uniform $\eta$ (Figures~\ref{fig:time_avg} and \ref{fig:rec_rate}).  The energy in the in-plane magnetic field decreases exponentially over a brief interval, while the vertical electric field $E_z$ surges and the peak current density $j_z$\footnote{We use lower-case $j_z$ to denote the current density estimated from the magnetic curl in a simulation, while ignoring the (typically small) displacement current, e.g. $j_z = (c/4\pi)(\bm{\nabla} \times \bm{B})_z$.} spikes.  This phase is followed by an oscillatory decay of the in-plane electromagnetic field, driven by slow azimuthal sloshing motions. These motions leave a clear imprint on the time profile of the hydromagnetic activity, with a period corresponding to the Alfv\'{e}nic crossing time halfway around the flux tube at $r=R/2$, $T \sim \pi (B_g/B_0) (R/c) \sim 7.8 t_c$ (Figures~\ref{fig:time_avg} and \ref{fig:rec_rate}).

In the non-rotating simulations, the in-plane electric field is less peaked in time, with $\langle E_\perp^2\rangle = B_g^2\langle v_\perp^2\rangle/c^2 = 25B_0^2\langle v_\perp^2\rangle/c^2$ reaching about 20\% of the change in magnetic energy and showing conspicuous oscillations thereafter.  Plasma heating is therefore driven by a combination of resistive heating during reconnection and damping of hydromagnetic motions. We note that the growth of plasma energy density $U^p = w\Gamma^2 - p$  
begins during the decay phase, but continues substantially into the later oscillatory phase as the plasma continues to be heated.  

\begin{figure}
    \centering
    \includegraphics[width=\linewidth]{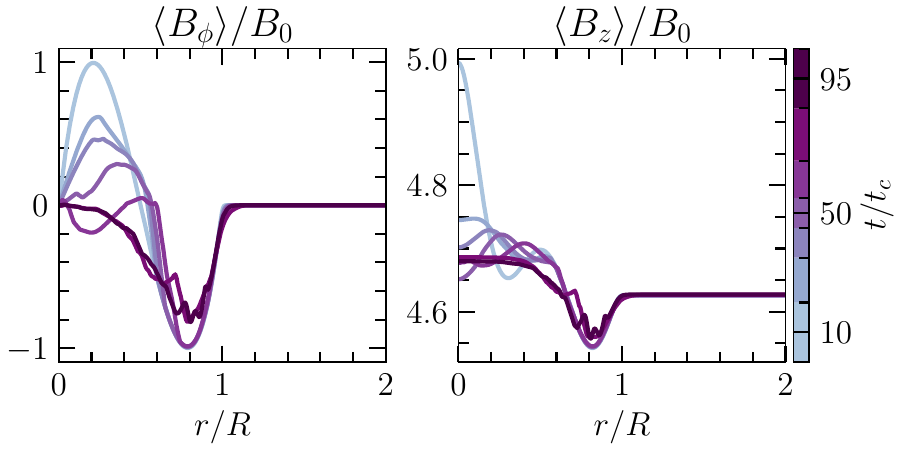}
    \caption{
    Evolution of the azimuthally-averaged magnetic field in \texttt{Rev2D}, showing separately the in-plane azimuthal and out-of-plane components $\langle B_\phi\rangle$ and $\langle B_z\rangle $. The central twist is reduced by a combination of reconnection and displacement in the first few snapshots. The times plotted are $t/t_c \in \{ 10, 30, 40, 50, 60, 75, 95 \}$, and each profile is averaged over a time interval $\pm 1.0 \, t/t_c$.
    }
    \label{fig:az_average}
\end{figure}

The dissipation of the inner current is revealed in the evolving profile of azimuthally-averaged toroidal field $\langle B_\phi\rangle$ (shown in Figure~\ref{fig:az_average}). At late times, the surviving twist matches the sign of the initial exterior twist, and vanishes inside the radius $\sim R/2$ where originally $B_\phi > 0$. The system does not fully relax to a state of uniform $J/B$.

\subsection{The Rate of Reconnection}

\begin{figure}
    \centering
\includegraphics[width=\linewidth]{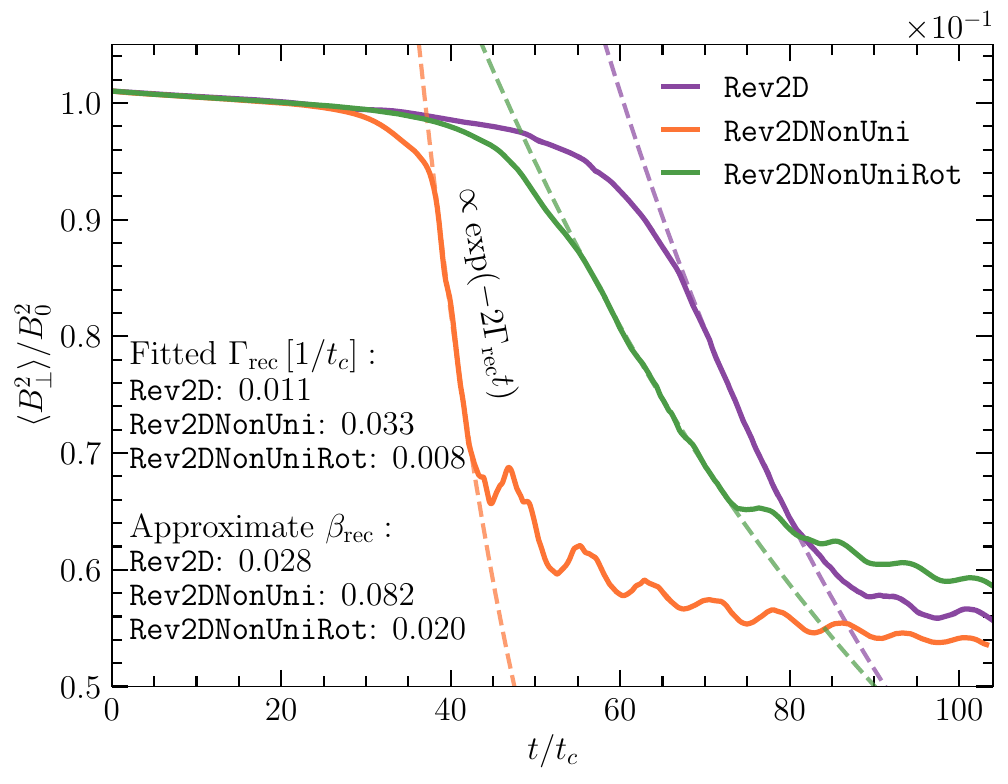}
    \caption{
    Reconnection of the in-plane magnetic field in simulations \texttt{Rev2D}, \texttt{Rev2DNonUni}, and \texttt{Rev2DNonUniRot} is concentrated within a narrow time interval, as shown by the time evolution of the domain-averaged in-plane magnetic field. An exponential decay, $\exp({-2\Gamma_{\rm rec} t})$, is fitted to the interval of peak reconnection (dashed line). The approximate reconnection rate is given by $\beta_{\rm rec} \sim (B_g/B_0) (R/c) \Gamma_{\rm rec}$, which varies between $\sim$ 0.02 and 0.1 across the simulations.
    }
    \label{fig:rec_rate}
\end{figure}

Reconnection of the toroidal field occurs in a single plasmoid-unstable current layer in the two non-rotating models. The onset of reconnection is pushed to progressively later times as one moves from simulation \texttt{Rev2DNonUni} (which spatially resolves charge-starved zones and is non-rotating), to the analogous rotating model \texttt{Rev2DNonUniRot} (where radial in-plane motions are slowed down by the Coriolis force), and finally to the uniform-$\eta$, non-rotating model \texttt{Rev2D}.  In the first case, the  dissipation of magnetic energy peaks over an interval $\Delta t_{\rm rec} \sim 5\,t_c$ centered around $t = 40\,t_c$; in the last case, the process expands to fill $\Delta t_{\rm rec} \sim 20\,t_c$ centered at $t = 75\,t_c$ (Figure~\ref{fig:rec_rate}).  

In contrast with simulations of reconnection with planar geometry, such as the well-known Harris sheet \citep{1962NCim...23..115H} (e.g. \citealt{2014PhRvL.113o5005G, 2014PhPl...21e6501C, 2015ApJ...806..167G, 2016MNRAS.462...48S, Werner2017ApJ, Hakobyan2019, Werner2021JPlPh, 2023ApJ...959..122C, 2025arXiv250320013G,2026MNRAS.546ag286B,Ripperda2026}), there is no ambiguity here about boundary conditions for the flow exiting the X-points. The flow, both into the reconnecting layer and out through the exhaust, is contained entirely within the computational domain.

The decay of the in-plane magnetic field may be fitted by an exponential
\citep{Bransgrove:2021}, $\langle B_\perp^2 \rangle \propto e^{-2\Gamma_{\rm rec} t}$
(Figure~\ref{fig:rec_rate}). This decay is expressed in terms of an in-plane
reconnection speed $v_{\rm rec}$ via
$\Gamma_{\rm rec} \sim v_{\rm rec}/R = \beta_{\rm rec} (B_0/B_g)(c/R)$.
From our fits, we infer $\beta_{\rm rec} \sim 0.1$ in
\texttt{Rev2DNonUni}, decreasing to $\beta_{\rm rec} \sim 0.03$ in
\texttt{Rev2D}. These values are consistent with prior Harris sheet
simulations in planar geometry, in the fast reconnection regime
\citep{2025arXiv250320013G, Sironi_2025, Ripperda2026}. The increased rate of
reconnection in \texttt{Rev2DNonUni} corresponds to faster bulk motions, which
result in the expulsion of flux from the original cylindrical domain (see
Figure~\ref{fig:jz_panel}, rightmost panel, second row). This correspondence is modified in the rotating case \texttt{Rev2DNonUniRot}, where other dissipation channels are evident (Figure~\ref{fig:jz_panel}). The decay of the in-plane magnetic field is much slower ($\beta_{\rm rec} \sim 0.02$) partly due to the effects of Alfv\'{e}nic shear \citep{2011PhPl...18g4501C, 2024ApJ...964..144P, 2024A&A...690A.389F}.

\begin{figure*}
    \centering
    \includegraphics[width=0.94\linewidth]{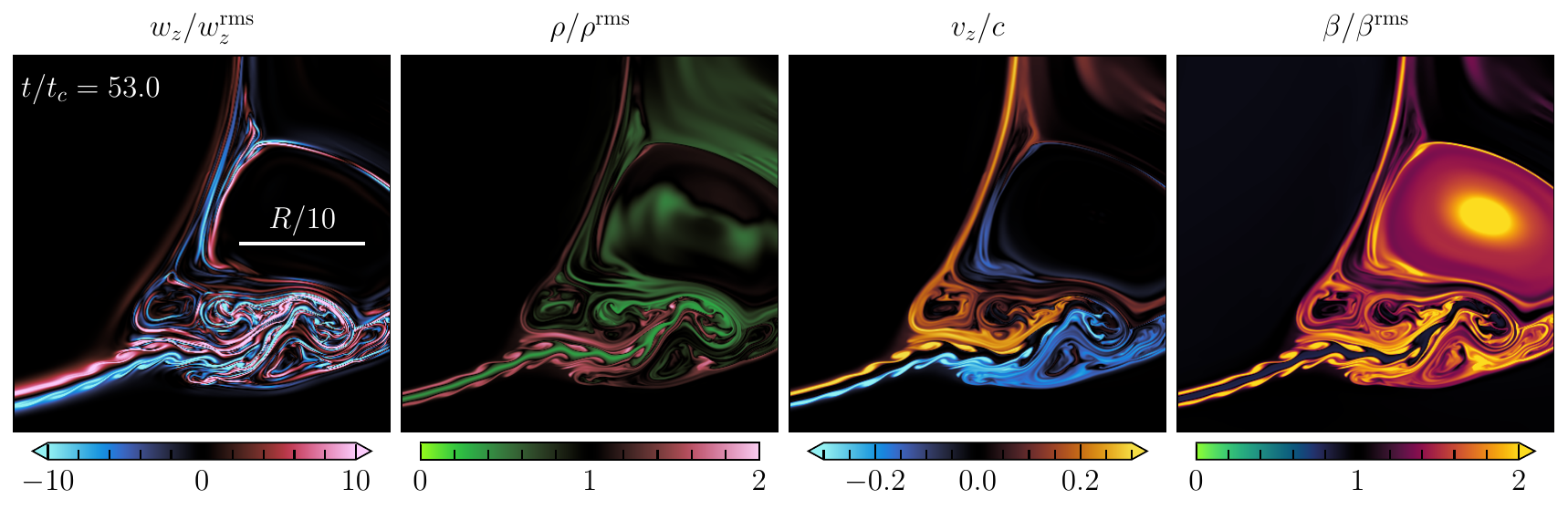}
    \caption{
    Zoom-in of the KH-unstable outflow from the reconnection layer after a plasmoid has been ejected into the downstream, in simulation \texttt{Rev2D}. Plotted are the vorticity, mass density, out-of-plane velocity and plasma $\beta$ parameter ($\beta = 8\pi p/B^2$). The first two and last quantities are normalized by their r.m.s. values in the domain $|x|, |y| \leq R$. The double vortex layer forms KH billows that help to seed the turbulent exhaust. These billows host large fluctuations (high-$k$ mode power) in $\rho$ and $p$.
    }
    \label{fig:khi}
\end{figure*}

\begin{figure*}
    \centering
    \includegraphics[width=0.94\linewidth]{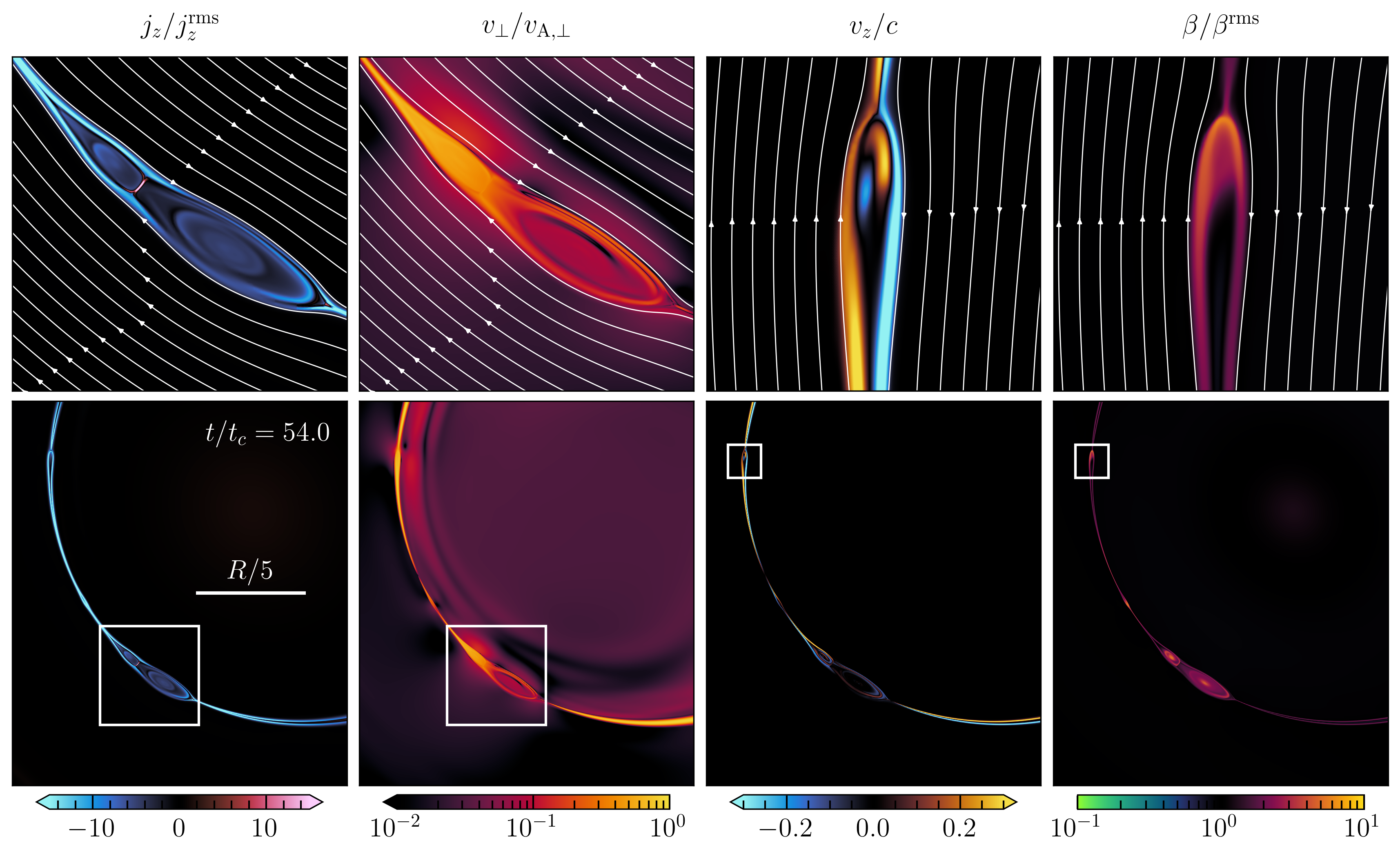}
    \caption{
    Zoom-in of the plasmoid-unstable reconnection layer in simulation \texttt{Rev2D}. Plotted are the out-of-plane current density, magnitude of the in-plane fluid velocity, out-of-plane fluid velocity and plasma $\beta$. The first and last quantities are normalized as in Figure~\ref{fig:khi}. Top row zooms further into the domains marked by a white rectangle in the bottom row.  Thin white lines show magnetic field lines.
    }
    \label{fig:plasmoid}
\end{figure*}

\subsection{Dynamics}\label{sec:dynamics}

The reconnecting current sheet forms two powerful outflows that 
rapidly become turbulent (Figure~\ref{fig:jz_panel};  see the expanded view in Figure~\ref{fig:khi}). This  
turbulent fluid is expelled and expands in volume as the central $I_z > 0$ current bundle dissipates. At any given time, the current sheet also segments into a handful of plasmoids that flow outward away from the central reconnection axis (see the expanded view in Figure~\ref{fig:plasmoid}). The plasmoids trap internal energy that is generated in neighboring X-points, thereby maintaining plasma $\beta = 8\pi p/B^2$ higher than the r.m.s. value of $\beta^{\rm rms} \sim 10^{-3}$. 

The reconnection outflows are Alfv\'{e}nic, with in-plane speed $v_{\perp} \simeq v_{{\rm A},\perp} = (B_0/B_g) v_{\rm A}$.  The in-plane and out-of-plane velocities are strongly sheared across the reconnection layer, with the latter switching its sign across the layer. A Kelvin-Helmholtz instability (KHI) develops in the horizontal flow (note that our 2D simulations cannot represent the KHI in the vertical flow). Figure~\ref{fig:khi} shows how linearly unstable KH billows collect in the turbulent exhaust. The ejection of a plasmoid triggers and potentially amplifies an enhanced level of turbulence downstream of the reconnecting current sheet.
Closely related structures are seen in simulations of merging flux tubes \citep{Ripperda2019, Velberg2025, Granier2026}.

The flow within the current sheet has a simple electromagnetic description, given the high magnetization and strong guide magnetic field. The outflows are bounded by vortex layers (Figure~\ref{fig:khi}) that are also sites of enhanced charge density. The linear charge density $\Sigma$ switches sign on either side of the outflow, as expected from the application of Gauss' law: $\Sigma = \pm E_\perp/4\pi = \pm (v_\perp/4\pi c) B_g$. To see this, we note that the in-plane electric field is oriented perpendicular to the outflow from an X-point, 
${\bm E}_\perp \simeq -{\bm v}_\perp\times (B_z \hat z)/c$. The vorticity is therefore directly proportional to the local charge density,
\begin{equation}\label{eq:vorticity}
    \omega_z = \hat z\cdot({\bm\nabla}\times{\bm v}_\perp) \simeq -c({\bm\nabla}\cdot{\bm E_\perp})/B_g.
\end{equation}

\begin{figure*}
    \centering
    \includegraphics[width=\linewidth]{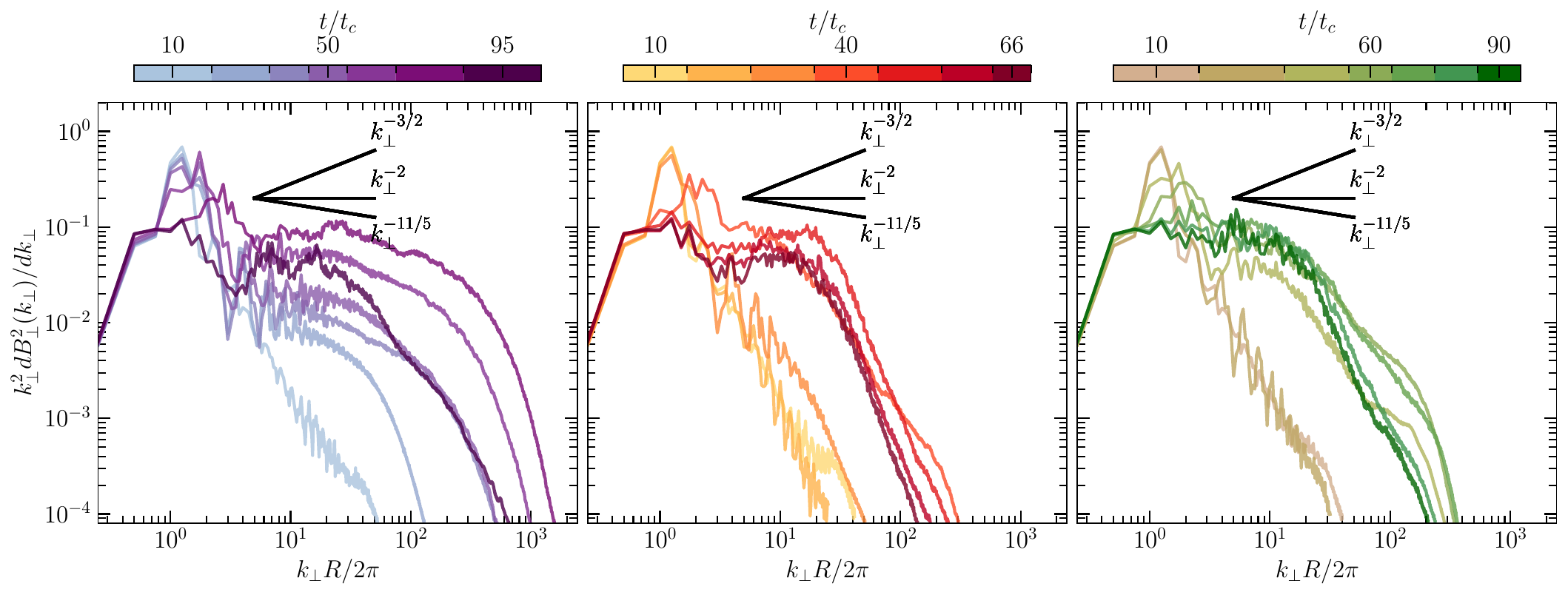}
    \caption{
    Time evolution of the perpendicular power spectrum of the in-plane magnetic field $B_\perp$, colored by $t/t_c$, showing the development of high-$k_{\perp}$ modes throughout the volume of the reconnecting flux domain in \texttt{Rev2D}, \texttt{Rev2DNonUni}, and \texttt{Rev2DNonUniRot}. Spectra are averaged over intervals $\pm 1t_c$. As reconnection begins, the spectrum develops a power-law component $dB_\perp^2(k_\perp) \propto k_\perp^{-2} dk_{\perp}$, which persists from $> 60 \, t_c$ at least to the end of the simulation. In \texttt{Rev2D} the snapshots match Figure~\ref{fig:az_average}; for \texttt{Rev2DNonUni} they are at times $t/t_c \in \{ 10,\, 20,\, 30,\, 40,\, 50,\, 60,\, 66 \}$, and for \texttt{Rev2DNonUniRot} at $t/t_c \in \{ 10,\, 30,\, 50,\, 60,\, 70,\, 80,\, 90 \}$.  
    }
    \label{fig:2D_spec}
\end{figure*}

As the turbulent exhaust develops, it becomes saturated with strong, small scale currents. To quantify the non-linear development of these electromagnetic fluctuations we use the one-dimensional, isotropically integrated Fourier power spectrum of the in-plane magnetic field,
\begin{equation}
    \frac{dB^2_\perp(k_\perp)}{dk_\perp} = \int k_\perp^2 d\Omega_{k_\perp} \bm{B}_\perp (k_\perp) \cdot \bm{B}_\perp^\dagger(k_\perp),
    \label{eq:spec_bperp}
\end{equation}
where $ \int k_\perp^2 d\Omega_{k_\perp}$ is the fixed-shell integral of the power at each $k_{\perp}$, such that, e.g., $\int dk_\perp  dB_{\perp}^2/dk_{\perp} = \left\langle B_{\perp}^2\right\rangle$ is satisfied (Parseval's theorem, \citealt{parseval1806memoire}). In Figure~\ref{fig:2D_spec}, we show the in-plane magnetic spectra for all three 2D simulations.

\begin{figure*}
    \centering
    \includegraphics[width=\linewidth]{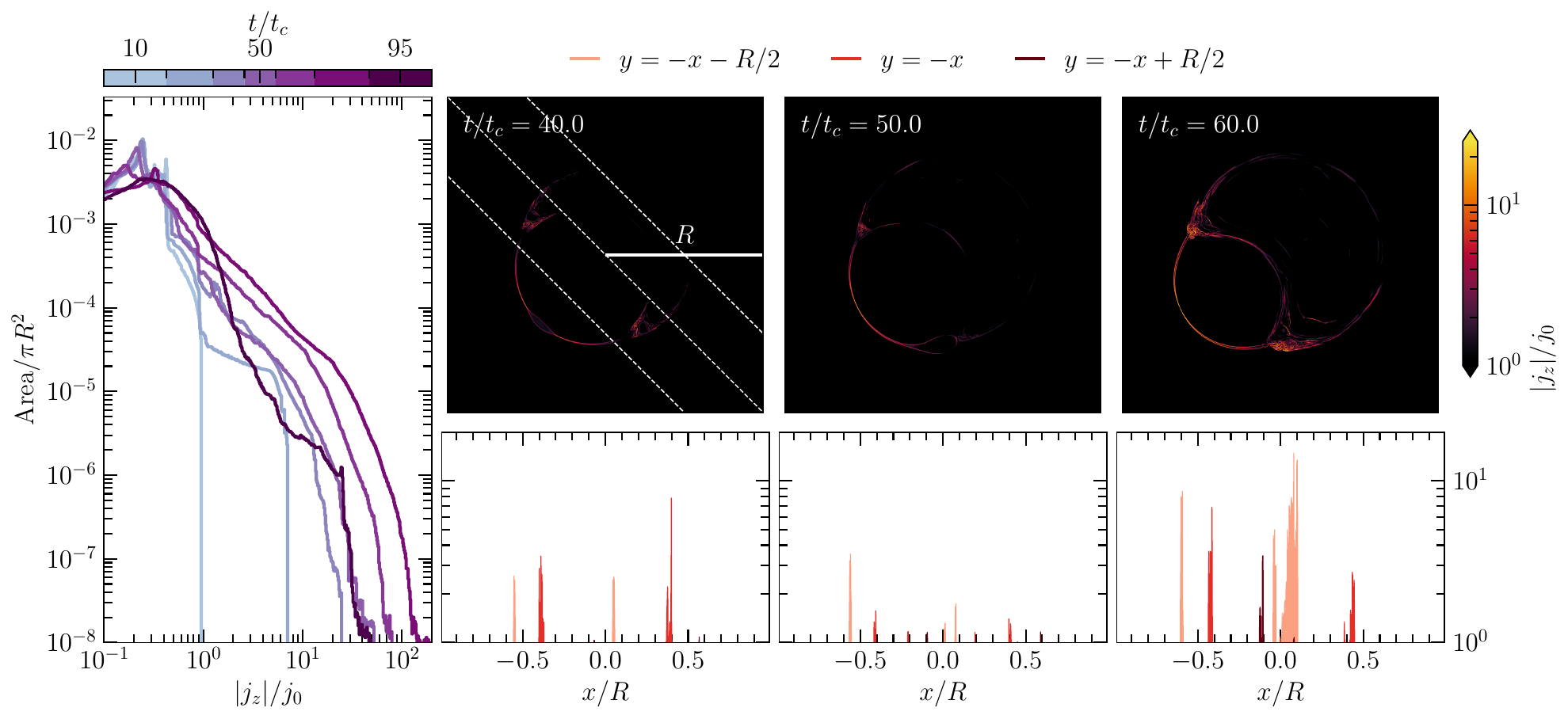}
    \caption{
    Evolution of the peak current density as internal reconnection develops in simulation \texttt{Rev2D}. Top panels: zones in which $j_z$ exceeds the peak background current density $j_0 = (c/4\pi) K_{\phi 1} B_0/R$ are marked in red. Each panel below shows three independent diagonal slices through the 2D $|j_z|$ contours above (dashed white lines). Colors pink, red and brown indicate slices increasing in $y$. Left panel: histogram of the area containing current density $|j_z|$, normalized to $j_0$. The histogram is averaged over time intervals $\pm 1.0 t/t_c$, and the times plotted match Figure~\ref{fig:az_average}.
    } \label{fig:current}
\end{figure*}

The magnetic spectrum develops a $dB_{\perp}^2/dk_{\perp} \sim k_\perp^{-2}$ profile for $k_\perp R/2\pi \gtrsim 5$ as reconnection sets in.
In our simulations, this spectrum emerges from the nonlinear development of the KHI in a thin reconnection outflow, as modified by secondary ohmic damping.
The same spectrum is measured in a network of strong discontinuities (Burgers turbulence), which may be either current sheets or regions of strong magnetic 
shear (e.g.,  \citealt{Beattie2025_supersonic_turbulence}) and are also seen to develop in simulations of
weak wave-wave turbulence \citep{Ripperda2021_FFR}.   There is equivalently a plateau in the current power spectrum, $dj_z^2/dk_{\perp} = k_\perp^2 dB_\perp^2 (k_\perp)/dk_\perp$, which broadens and flattens during the peak of reconnection. Likewise, as shown in Figure~\ref{fig:current}, when the power shifts to larger $k_\perp$, the strong small scale currents occupy a larger portion of the initial flux tube domain. 

\begin{figure}
    \centering
    \includegraphics[width=0.83\linewidth]{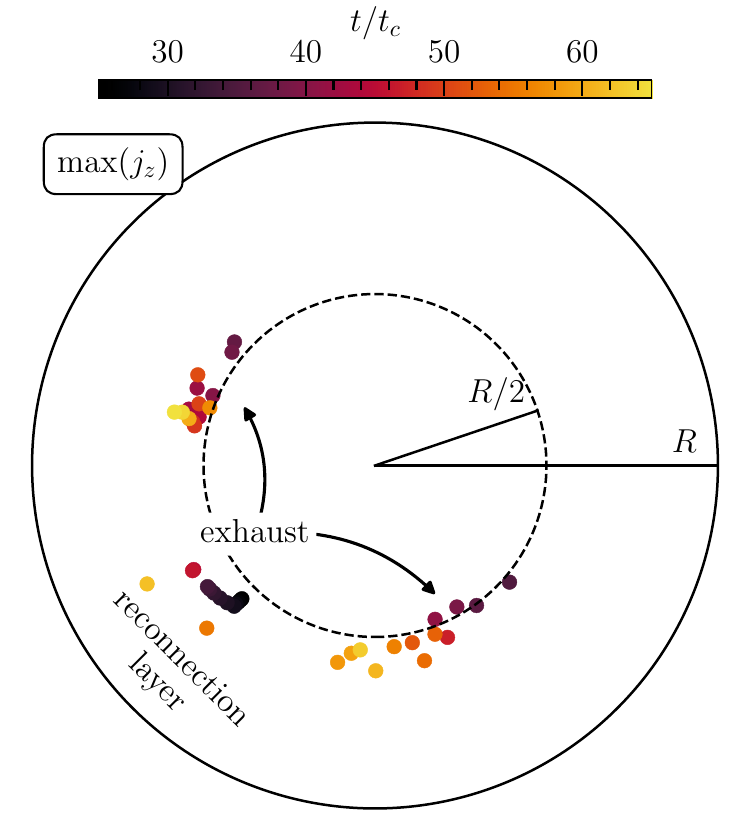}
    \caption{
    The evolution of the location of maximum $j_z = (c/4\pi) (\bm{\nabla}\times\bm{B})_z$ in simulation \texttt{Rev2D}. At early times, the position of this maximum tracks the plasmoid-unstable reconnection layer, while occasionally switching to the far turbulent exhaust. At late times, the maximum is concentrated in the exhaust. Asymmetry in the exhausts is driven by the ejection of large plasmoids forming by coalescence in the sheet (preferentially in the bottom-right exhaust; see Figure~\ref{fig:jz_panel}). The surface across which $B_\phi$ reverses sign in the initial condition is indicated by a dashed black circle, while the outer boundary of the flux tube is indicated by a solid black circle. }
    \label{fig:jz_loc}
\end{figure}

\begin{figure}
    \centering
    \includegraphics[width=\linewidth]{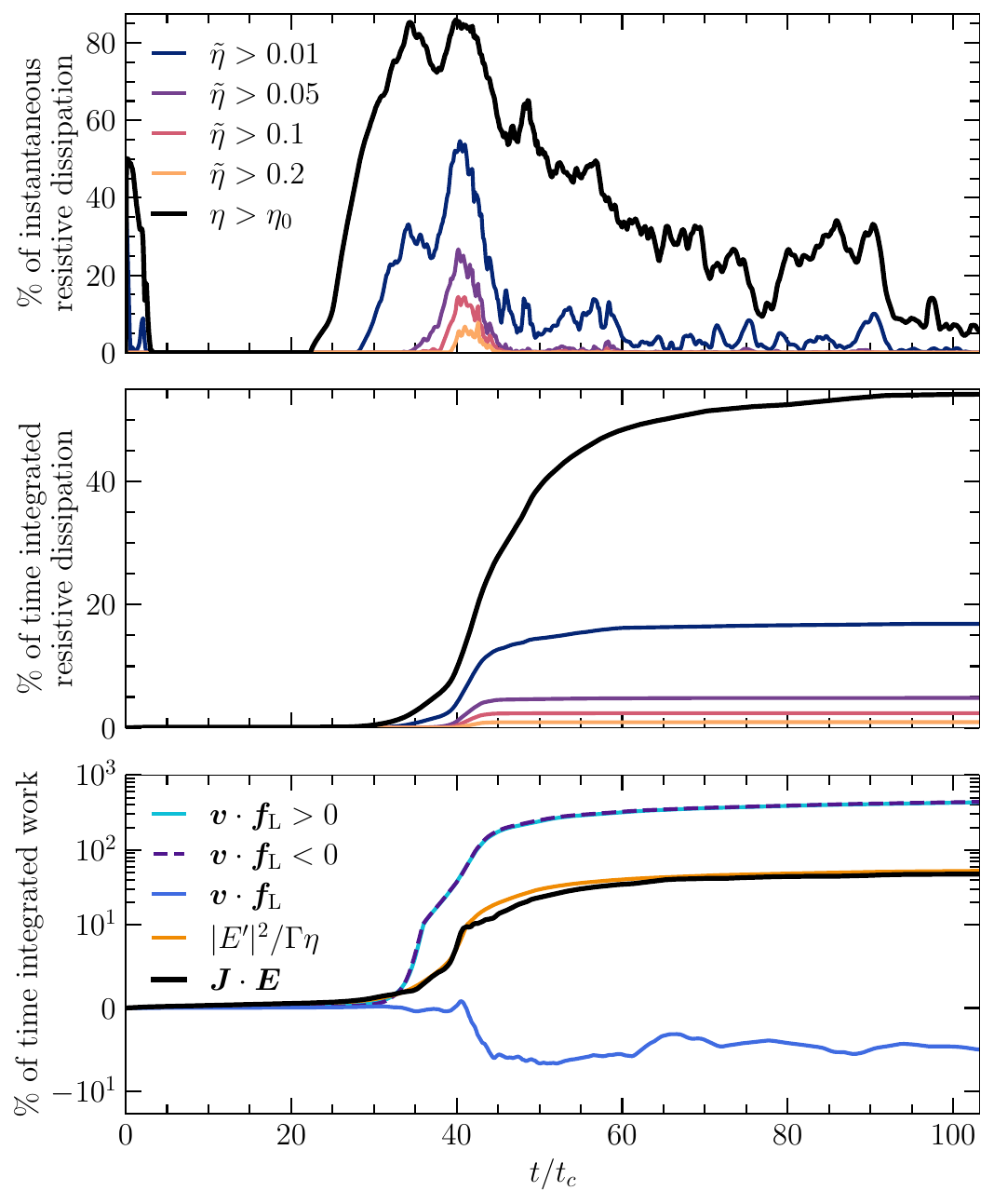}
    \caption{
    Time evolution in \texttt{Rev2DNonUni} of the percent of instantaneous \eqref{eq:instant_res_diss} and time integrated \eqref{eq:total_res_dis} resistive dissipation (top and middle panel), and percent of the time integrated work \eqref{eq:total_dis_comp} broken into components (bottom panel). In the top two panels, the contribution to the resistive dissipation for charge-starved layers above indicated thresholds are shown \eqref{eq:eta_tilde}.
    }
    \label{fig:threshold_work}
\end{figure}

\begin{figure}
    \centering
    \includegraphics[width=\linewidth]{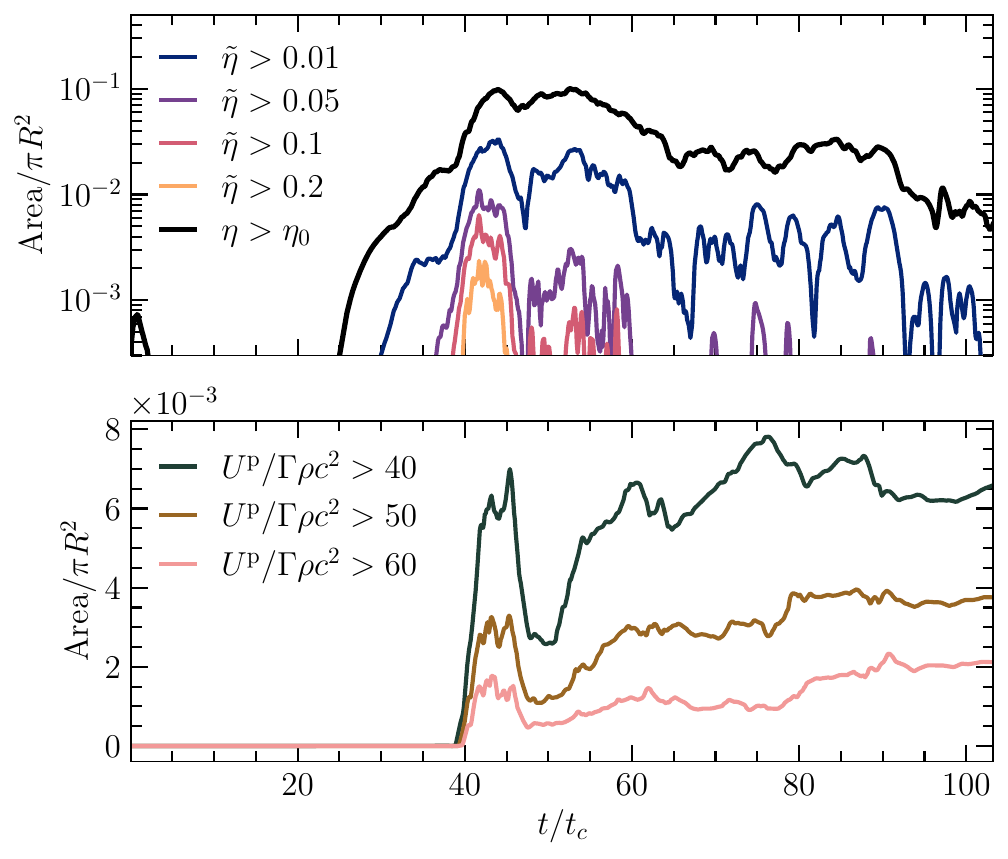}
    \caption{
    Time evolution in \texttt{Rev2DNonUni} of the area occupied by regions of charge starvation above specified thresholds \eqref{eq:eta_tilde} (top) and regions of large plasma energy relative to lab frame mass energy density (bottom).}
    \label{fig:threshold_areas}
\end{figure}

\begin{figure}
    \centering
    \includegraphics[width=\linewidth]{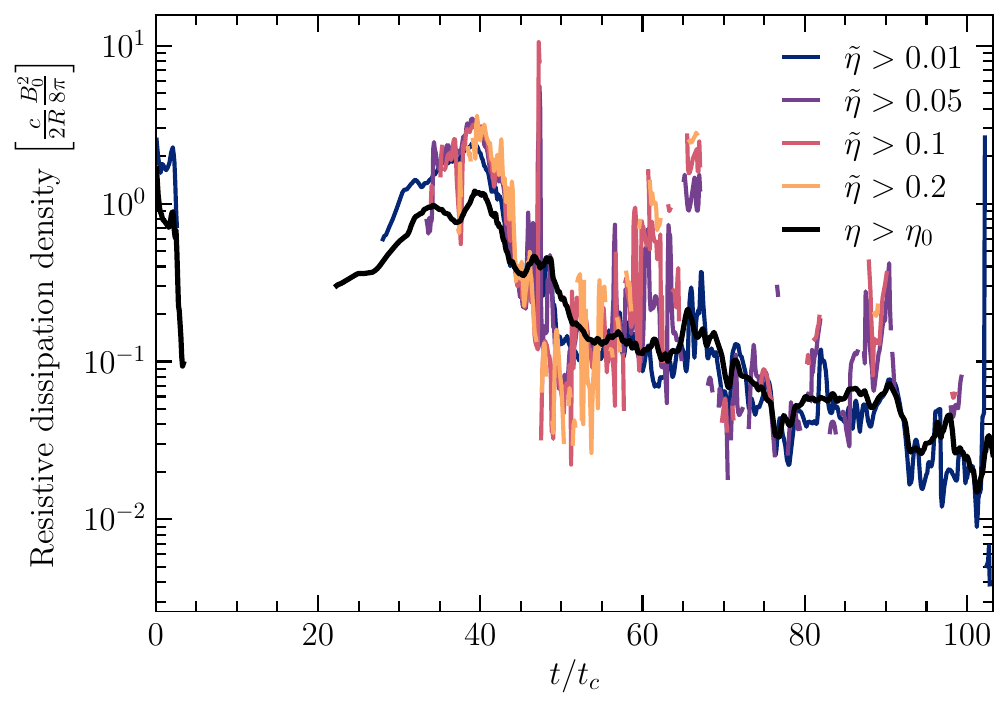}
    \caption{
    The volumetric rate of energy dissipation in zones with different levels of charge starvation (colored lines) for \texttt{Rev2DNonUni}.  One observes that the dissipation rate plateaus for $\tilde\eta > 0.01$.  For reference, the black line shows the volumetric average in all zones with $\eta > \eta_0$.
    }
    \label{fig:dissipation_density}
\end{figure}

\begin{figure*}
    \centering
    \includegraphics[width=0.94\linewidth]{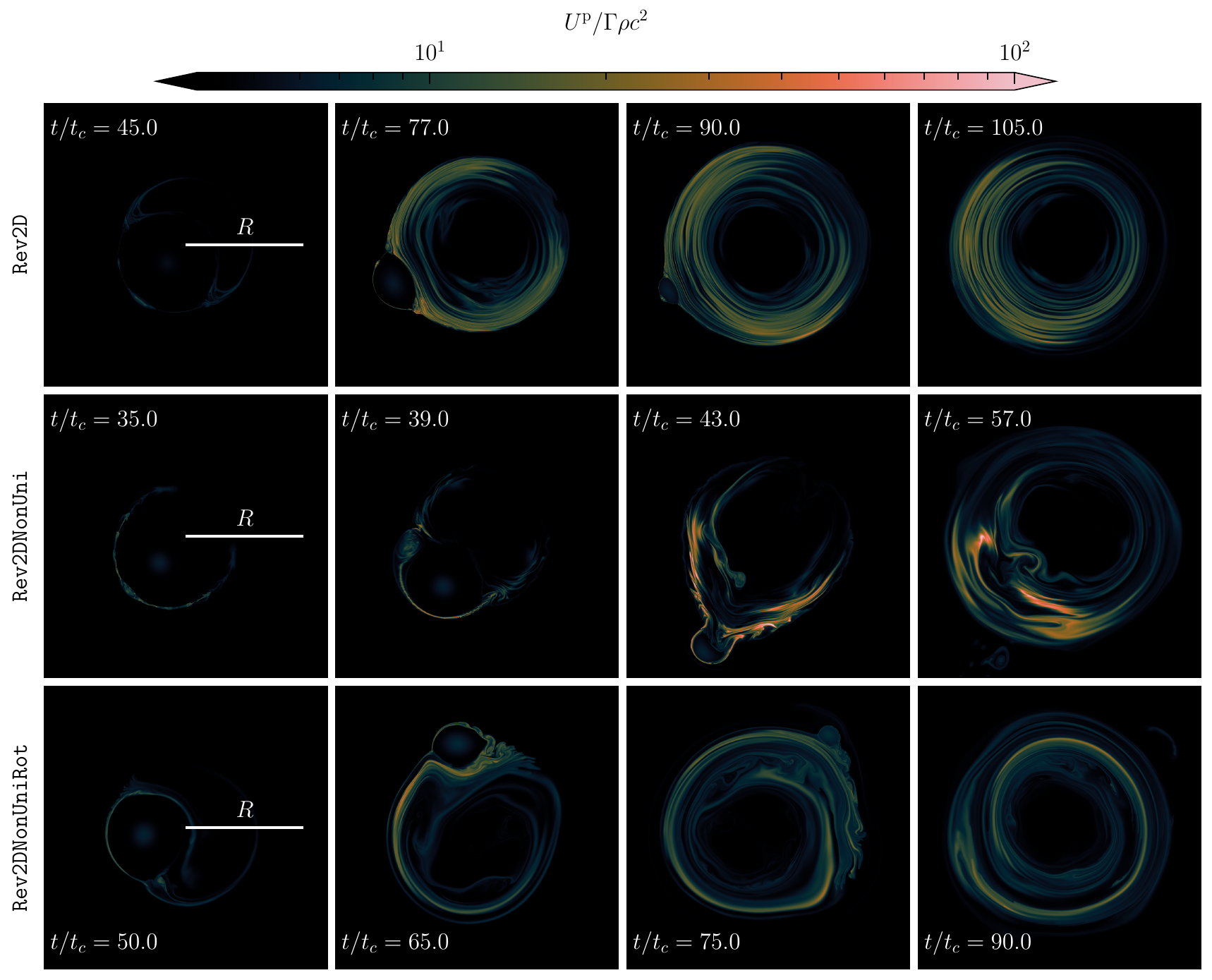}
    \caption{
    Time evolution of ratio between the plasma energy $U^{\rm p} = \rho h \Gamma^2 c^2  - p$ and lab frame mass density energy, $\Gamma \rho c^2$, for simulations \texttt{Rev2D}, \texttt{Rev2DNonUni}, and \texttt{Rev2DNonUniRot}. 
    }
    \label{fig:Uintpanel}
\end{figure*}

\begin{figure*}
    \centering
    \includegraphics[width=0.94\linewidth]{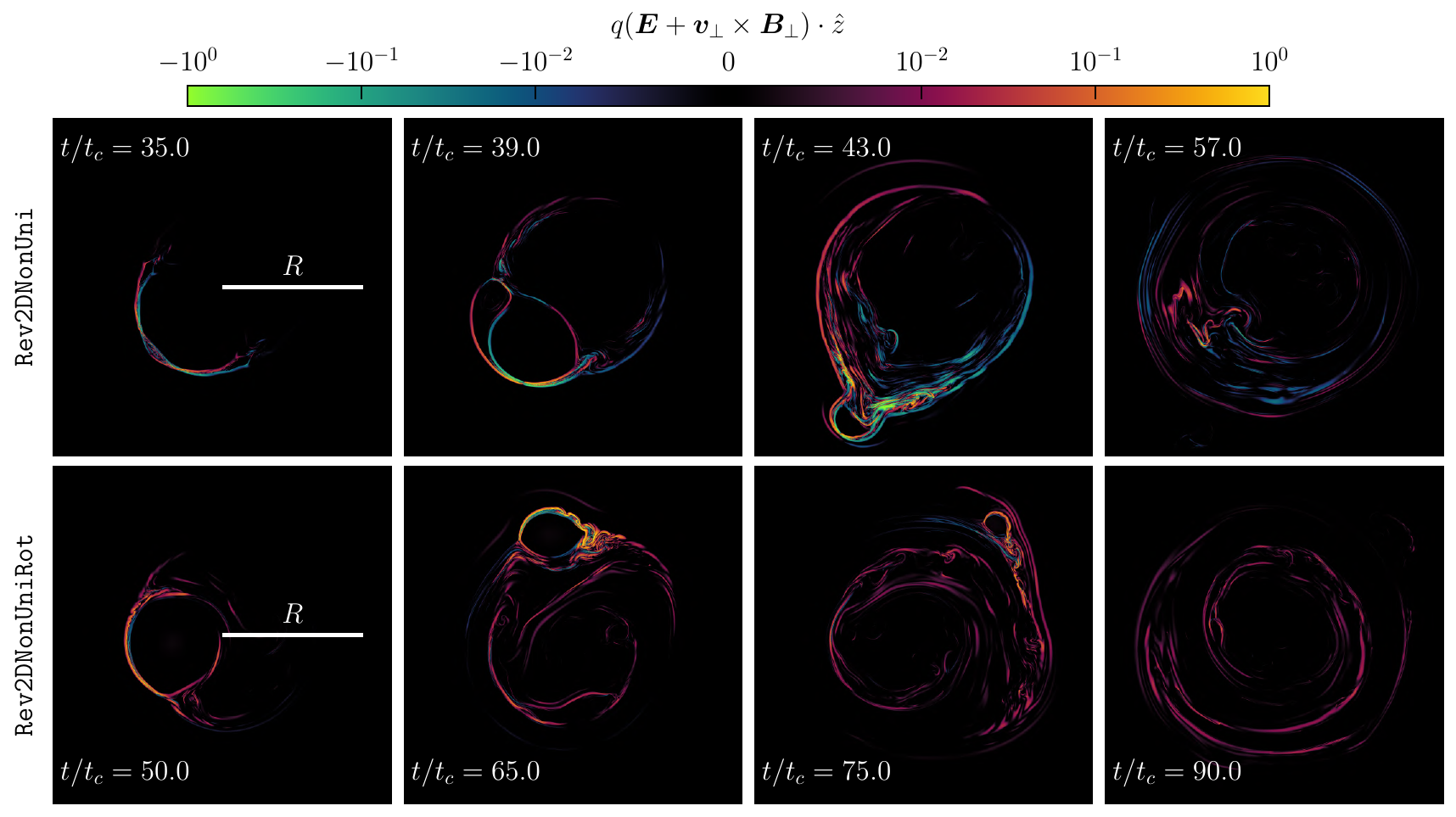}
    \caption{
    Net out-of-plane Lorentz force $({\bm J}_\perp\times{\bm B}_\perp)\cdot\hat z + qE_z \sim q (\bm{E} + \bm{v}_\perp \times \bm{B}_\perp) \cdot \hat{z}$ in the two simulations with modified resistivity law \eqref{eq:etamod} and resolved charge-starved zones. The rotating model \texttt{Rev2DNonUniRot} shows a strongly asymmetric acceleration along the guide magnetic field. 
    }
    \label{fig:qEz}
\end{figure*}

\begin{figure*}
    \centering
    \includegraphics[width=0.94\linewidth]{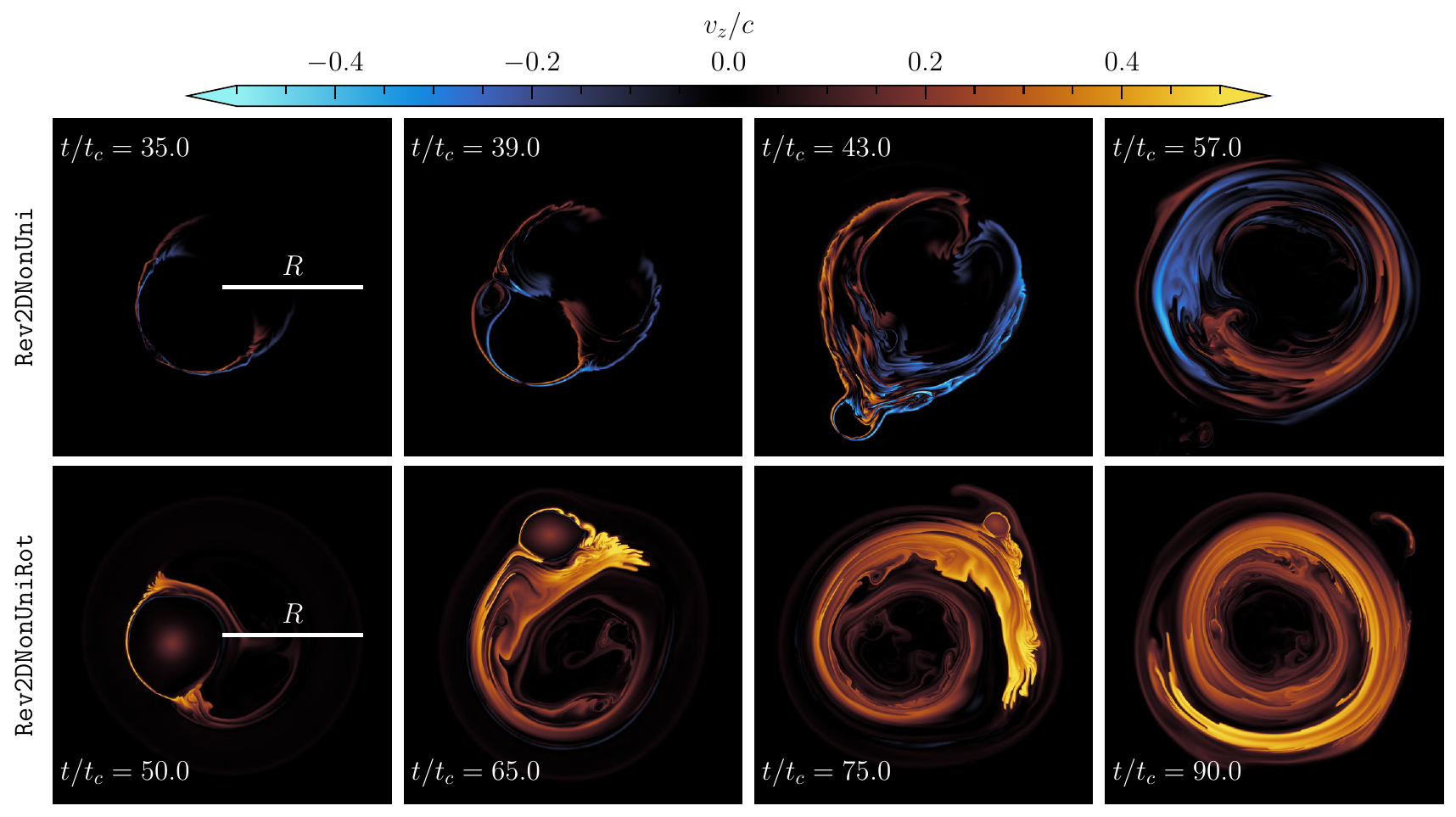}
    \caption{
    Developing out-of-plane velocity in the two simulations with modified resistivity law \eqref{eq:etamod}. The fluid is accelerated preferentially in the reconnecting current sheet and then advected around the plane along with the connected guide magnetic field. Mixing triggered by the KHI in the outflow brings together field lines that have experienced strong non-ideal electric field with those that have not. 
    }
    \label{fig:vz}
\end{figure*}

\subsection{3D Dynamics}\label{sec:3d_dyn}

The validity of the planar flow approximation made in our 2D simulations has been tested by extending the field structure uniformly into a third dimension along the flux tube. The dynamics of the 3D, ideal MHD simulations described in Appendix~\ref{app:3D} is similar to that observed in model \texttt{Rev2D} for identical guide field strength. Reconnection drives motions in the $xy$-plane, and almost no structure is generated in the vertical direction. The current sheet reconnects and the two resulting outflows become turbulent. Nearly the entire inner twist is dissipated through reconnection, as is seen in 2D.  When the guide field is reduced to
$B_g = 2.25 B_0$, the flux tube becomes kink unstable, as the fastest-growing kink mode outpaces the reconnection that is triggered by the local perturbation ($\Gamma_{\rm kink}/\Gamma_{\rm rec} \sim 3$ from Equation~\eqref{eq:rec_vs_kink_rate} with $\beta_{\rm rec} \sim 0.03$). The dynamics are then truly 3D. At early times, the growing kink mode forms lobe-like structures that are periodic in the vertical coordinate. These lobes are contained within radius $r < R/2$, that is, within the inner zone with $I_z > 0$. Reconnection is ignited on the radial boundary of the lobes, forming current sheets that are displaced over a range of radius. As reconnection dissipates the inner twist, it generates structure with both high $k_\perp$ and $k_z$.  This structure first develops in the $I_z > 0$ core of the flux tube, similarly to the $m=1$
mode in a flux rope with a uniform sign of current (e.g. \citealt{Davelaar2020}).

For comparison, \cite{Striani2016} have studied spontaneous 2D and 3D reconnection in a similar configuration with reversing magnetic field but a weaker guide field ($B_z$ equal to twice peak $B_\phi$) that decays as $B_z(r) \sim r^{-2}$ at large radius, with application to astrophysical jets. In the absence of a localized perturbation, these 2D simulations (maximum resolution $4096^2$) show the spontaneous development of multiple magnetic islands. The $m=1$ kink mode develops strongly in 3D, as expected from our analysis.

\subsection{Intermittency of Dissipation}

The highest current density encountered in these simulations is localized in the large-scale current sheet and the surrounding turbulent, reconnection-driven outflow. As reconnection proceeds, the position of peak current density shifts away from the reconnection axis to the exhaust zone (Figure~\ref{fig:jz_loc}). The area of the 2D plane that is filled with $|j_z| = (c/4\pi)|\bnabla\times{\bm B}|_z$ higher than the peak background current density $j_0$ is shown for simulation \texttt{Rev2D} in the left panel of Figure~\ref{fig:current}.  At the peak of dissipation, this distribution shows a tail, comprising area $\sim 10^{-3} j_0|j_z|^{-1}\pi R^2$ and extending to a current density $|j_z|/j_0 \sim 30$.  

Although the areal coverage is relatively small ($\sim 10^{-3} j_0|j_z|^{-1}\pi R^2$), a slice at constant coordinate $x + y$ through the simulation domain reveals significant isolated dissipation spikes (see the lower panel of Figure~\ref{fig:current}).  These slices correspond to possible motions of a fixed line of sight
through the plasma in the case where the magnetic field is tied to a rotating object;  Section~\ref{sec:starve} discusses potential implications for the time-variable radio emission of a magnetar.

Consider now the connection between charge starvation and ${\bm E}\cdot{\bm B}$ dissipation in model $\texttt{Rev2DNonUni}$, which resolves charge-starved zones. The transfer of electromagnetic energy to plasma kinetic and thermal energy is described by Poynting’s theorem \citep{Poynting}
\begin{equation}
    \frac{\partial }{\partial t}U^{\rm EM} + \bnabla\cdot\bm{S} = - \bm{J}\cdot\bm{E},
\end{equation}
where $U^{\rm EM}$ is the electromagnetic energy density and $\bm{S} = c\bm{E}\times\bm{B}/4\pi$ is the Poynting flux. The source term $-\bm{J}\cdot\bm{E}$ represents the local rate at which electromagnetic energy is transferred between the plasma and the electromagnetic field. For a relativistic resistive plasma, this energy transfer is decomposed as
\begin{equation}
    \bm{J}\cdot\bm{E} =
    \underbrace{
    \bm{v} \cdot \left( q \bm{E} + \frac{\bm{J}\times\bm{B}}{c} \right)
    }_{\substack{\text{work done by} \\ \text{the Lorentz force}}}
    +
    \underbrace{
    \frac{|E^\prime|^2}{\Gamma \eta}
    }_{\substack{\text{resistive} \\ \text{dissipation}}}.\label{eq:dissipation}
\end{equation}
The first term describes reversible energy exchange between the electromagnetic field and the bulk plasma motion through the Lorentz force, $\bm{f}_{\rm L} = q \bm{E} + \bm{J}\times\bm{B}/c$. The second term is strictly positive and corresponds to irreversible conversion of electromagnetic energy into thermal energy via resistive dissipation, mediated by the fluid-frame electric field,
\begin{equation}
    \bm{E}^\prime = \Gamma \left[ \bm{E} + \frac{\bm{v}\times\bm{B}}{c}
    - \frac{1}{c^2}\frac{\Gamma}{\Gamma + 1} (\bm{v}\cdot\bm{E}) \bm{v} \right].
\end{equation}
Using $\bm{E}^\prime = \eta \bm{J}^\prime$, the resistive term may also be written in a form reminiscent of the Newtonian limit as $\eta|\bm{J}^\prime|^2/\Gamma$, where $\bm{J}'$ is the fluid-frame current density.

The contribution to the net dissipation from zones with different levels of enhanced resistivity is shown in Figure~\ref{fig:threshold_work}. We define the dimensionless resistivity threshold
\begin{equation}
    \tilde{\eta} \equiv \left(\frac{B_g}{B_0} \right)
    \left( \frac{|\bm{E}^\star|}{B_0} \right)
    \left( \frac{\rho_0}{\Gamma \rho} \right),
\label{eq:eta_tilde}
\end{equation}
or, equivalently, 
\begin{equation}
    \tilde{\eta} = \left(\frac{B_g}{B_0}\right)
    \left(\frac{\eta}{4\pi}\right)
    \left(\frac{2R}{\Delta_0}\right)
    \frac{1}{t_c};
\end{equation}
see Equation~\eqref{eq:etamod0}.
The plasma density is suppressed in the current sheet and we have $\left(B_g/B_0 \right) \left( |\bm{E}^\star|/B_0 \right) \sim \beta_{\rm rec}$ near an X-point,
implying that $\tilde{\eta} \gtrsim \beta_{\rm rec}$ in reconnection zones. We quantify the contribution of these regions to dissipation by defining the fractional 
resistive dissipation above a threshold $\tilde{\eta}_{\rm th}$,
\begin{equation}
    \left( 
    \begin{array}{c}
    \text{\% of instantaneous} \\
    \text{resistive dissipation}
    \end{array}
    \right)
    = \frac{\displaystyle \int_{\mathcal{D}_{\rm th}} dxdy\,W_\eta}
         {\displaystyle \int_{\mathcal{D}} dxdy\,W_\eta},
    \label{eq:instant_res_diss}
\end{equation}
where $\mathcal{D}_{\rm th} = \{ (x,y) | \tilde{\eta} > \tilde{\eta}_{\rm th} \}$, $\mathcal{D}$ is the entire computational domain, and $W_\eta = |E'|^2/\Gamma \eta$. Similarly, for the time-integrated resistive dissipation,
\begin{equation}
    \left( 
    \begin{array}{c}
    \text{\% of time integrated} \\
    \text{resistive dissipation}
    \end{array}
    \right)
    = \frac{\displaystyle \int_0^t dt\int_{\mathcal{D}_{\rm th}} dxdy\,W_\eta}
         {\displaystyle \int_0^{t_f} dt\int_{\mathcal{D}} dxdy\,W_\eta},
     \label{eq:total_res_dis}
\end{equation}
where $t_f$ is the final time of the simulation. The contributions by individual components of the work are calculated over the entire domain and measured relative to the total time integrated work
\begin{equation}
\left( 
\begin{array}{c}
\text{\% of time} \\
\text{integrated work}
\end{array}
\right)
= \frac{\displaystyle \int_0^t dt\int_{\mathcal{D}} dxdy\,W_a}
     {\displaystyle \int_0^{t_f} dt\int_{\mathcal{D}}dxdy\,(\bm{J}\cdot\bm{E})}.
     \label{eq:total_dis_comp}
\end{equation}
for component $W_a$, e.g. $W_a = \bm{v}\cdot \bm{f}_{\rm L} >0$.

During the peak of reconnection, charge-starved zones contribute 20\%-55\% of the instantaneous resistive dissipation (see the curve $\tilde\eta > 0.01$ in Figure~\ref{fig:threshold_work}). Following the peak of reconnection, resistive dissipation is dominated by Ohmic processes in an expanding network of thin circular current sheets. Secondary tearing and charge starvation effects are not observed in this outer zone with the resolution achieved here.

The relative work done by the Lorentz force is shown in the bottom panel of Figure~\ref{fig:threshold_work}. The contributions from the two signs of the Lorentz force, $\bm{v} \cdot \bm{f}_{\rm L} < 0$ and $\bm{v} \cdot \bm{f}_{\rm L} > 0$, nearly cancel as the medium oscillates and energy is repeatedly exchanged  between the electromagnetic field and the bulk kinetic motions. However, the net work done by the Lorentz force remains negative and contributes $\sim 10\%$ of the total dissipation. Dissipation following the peak of reconnection is dominated by slower, resistive effects.

In spite of contributing significantly to the resistive dissipation during reconnection, the charge-starved regions occupy a small fraction of the domain, 0.1\%-3\% of the flux tube's initial area, as shown in Figure~\ref{fig:threshold_areas}.  The volumetric rate of energy dissipation plateaus in zones with $\tilde\eta > 0.01$, extending up to $0.2$, as is shown in Figure~\ref{fig:dissipation_density}.  Zones with higher resistivity are effectively more depleted in charge carriers, and therefore show a higher dissipation rate per unit mass. Indeed, the introduction of the resistivity law (\ref{eq:etamod}) produces zones with greatly enhanced specific rates of dissipation, e.g., compare simulations \texttt{Rev2D} and \texttt{Rev2DNonUni} in Figure~\ref{fig:Uintpanel}.

\section{High-Wavenumber Currents and Charge Starvation}\label{sec:starve}

The outflow from the X-point produces a nested sequence of azimuthal current layers of varying sign (Figure~\ref{fig:jz_panel}). At any given time, a small fraction of the flux tube volume is so strongly sheared that $|j_z|$ exceeds the peak background current density $j_0$ (Figure~\ref{fig:current}). These zones are prime sites for relativistic particle acceleration if the background electron (and positron) density $n_0$ is not much larger than what is required to sustain the equilibrium current, i.e., the multiplicity ${\cal M} = n_0 e c/j_{0}$ is not much larger than unity. Then the strongest current sheets become charge-starved, and the particles within them gain energy until secondary pair creation sets in (e.g. \citealt{Hakobyan2019}).

To understand the implications for particle heating, we refer to a simpler model in which dilution of the particle density $n$ is driven by charge starvation. The in-plane magnetic field is taken to have Cartesian symmetry and to reverse sign at $x = 0$, with $B_y(x,t) \rightarrow +/- B_{y\infty}(t)$ at large positive/negative $x$. The strong guide magnetic field $B_g$ is sheared over a length\footnote{In a force-free magnetosphere, the quantity $L$ is invariant along a diverging magnetic flux bundle, because the ratio $J/B$ is also invariant. For example, in the open flux bundle of a rotating neutron star, this length is the size of the corotating magnetosphere, $L \simeq B/4\pi|\rho_{\rm co}| = c/2\Omega$, where $\rho_{\rm co} \sim -\Omega B/2\pi c$ is the corotation charge density and $\Omega$ the angular frequency of rotation.}  $L = B_g/|{\bm\nabla}\times{\bm B}|$. The net density of $e^\pm$ is $n \simeq n_\infty = {\cal M}|\bnabla\times{\bm B}|/4\pi e$, where $e$ is the magnitude of the electron charge. A maximum thickness $2x_s = 2x_{s,c}$ for a dissipative current sheet is obtained by setting $J_z$ equal to the limiting conduction current $en_\infty c$, 
\begin{align}\label{eq:xs}
    B_y &=  4\pi n_\infty e x; \quad(|x| < x_s)\nonumber\\
    x_{s,c} &= {B_{y\infty}\over 4\pi n_\infty e} = {\cal M}^{-1}{B_{y\infty}\over B_g}\,L.
\end{align}
We consider a situation where the particle density decreases with time; a concrete model of particle depletion is described in Section~\ref{sec:rho_in_charge_starved_zones}. Given that this happens adiabatically, the current remains marginally charge-starved and the parallel electric field $E_\parallel = \hat{B}\cdot {\bm E}$ satisfies $\partial E_\parallel/\partial x \simeq -c^{-1}\partial B_y/\partial t$.  The energy gained per (relativistic) particle $\Delta{\cal E} = \int dt e E_\parallel $ is obtained by integrating Faraday's law in both $t$ and $x$,
\begin{align}
    {\Delta{\cal E}\over m_ec^2} = \left({x\over x_s}\right)^2
    {B_{y\infty}^2\over 8\pi n_\infty m_ec^2}
    = {\sigma_\infty\over 2}\left({x\over x_s}\right)^2, \quad (|x| < x_s).
\end{align}
The heated particles become relativistic when the magnetization $\sigma_\infty = B_{y\infty}^2/4\pi n_\infty m_ec^2$ is large.
\begin{figure*} 
    \centering
    \includegraphics[width=\linewidth]{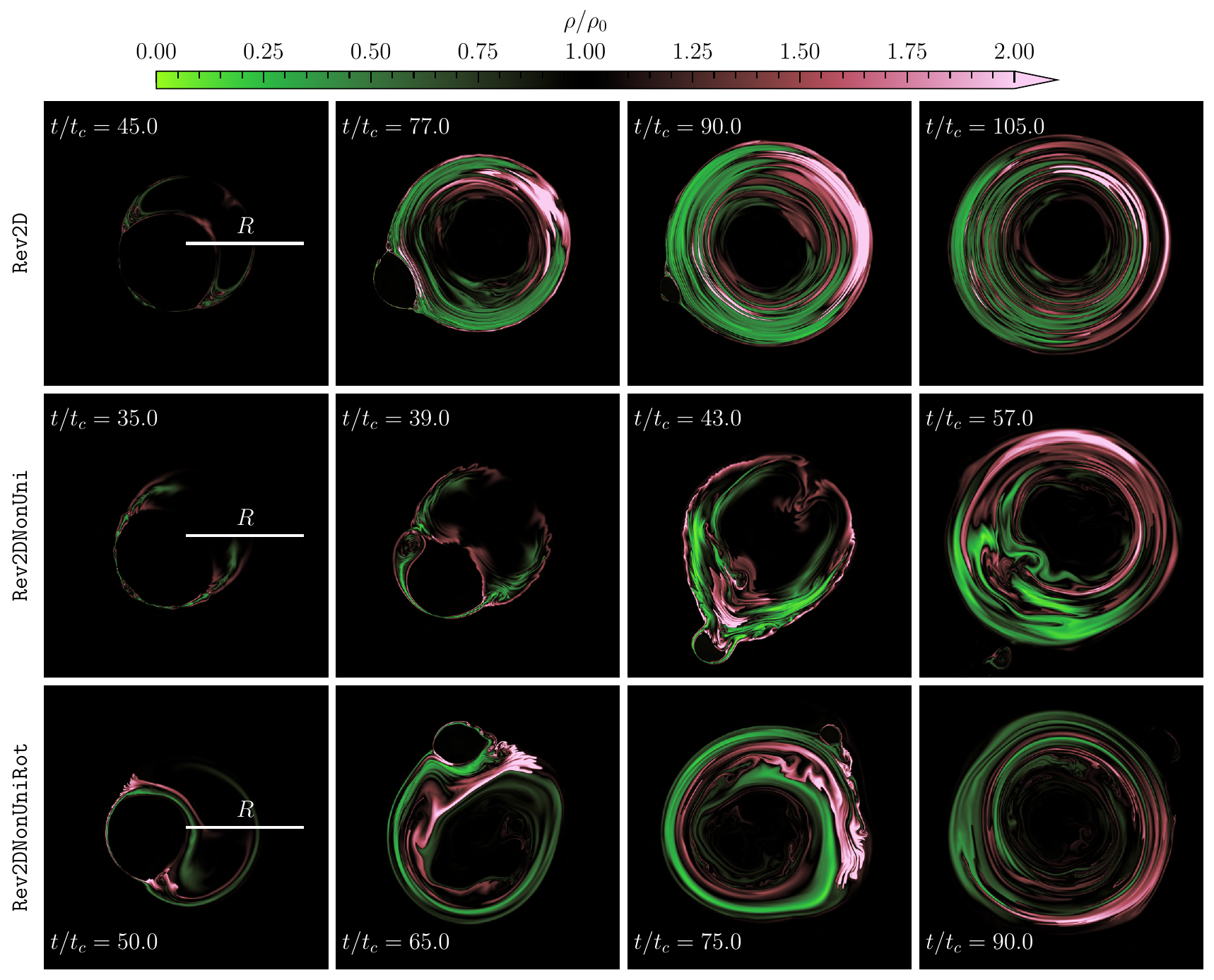} 
    \caption{ Time evolution of ratio between the fluid frame mass density and the initial uniform density in simulations \texttt{Rev2D}, \texttt{Rev2DNonUni}, and \texttt{Rev2DNonUniRot}.  } \label{fig:rho} 
\end{figure*} 
In the simulations presented here, the change in magnetization driven by dissipation of the in-plane magnetic field is $-\Delta\sigma = \mathcal{O}(B_\perp/B_g)^2\sigma$ and therefore is only mildly relativistic. A fraction of this dissipated energy is converted to bulk kinetic energy along the guide magnetic field (see the panel $\langle v_z^2\rangle/c^2$ in Figure~\ref{fig:time_avg}). 

The outflows from reconnecting current sheets show a concentration of non-ideal $E_z$ (see Figure~\ref{fig:qEz}); the out-of-plane acceleration is further concentrated in the vortex layers bounding the outflows, where the charge density $q$ is strongly enhanced.  Fluid with large vertical momentum is advected away from the current sheet and then more broadly distributed across the simulation domain (see Figure~\ref{fig:vz}).

In the magnetosphere of a quiescent magnetar or radio pulsar, the effective 
magnetization $(B_\perp/B)^2\sigma$ is typically much larger than unity, especially in parts of the outer magnetosphere that may host radio emission. In the presence of a strong guide magnetic field, as considered here, the momentum distribution of the heated particles becomes relativistic and tends to align with the overall magnetic field configuration \citep{Comisso2020}. The strongly heated zones are then a promising source of temporally variable electromagnetic emission, which is modulated as seen by a fixed observer by bulk rotation or long-wavelength hydromagnetic oscillations \citep{T06, Sobacchi2023}. 

\subsection{Particle Depletion in Charge-Starved Zones}\label{sec:rho_in_charge_starved_zones}

In the ideal MHD limit, time-dependent planar flow conserves the ratio $n/B_z$ of particle number to magnetic flux. The drift of charge across the magnetic field becomes significant in zones of high current density, especially where the particle density is too low to supply the current, $enc < |{\bm J}|$.  The resistivity is further enhanced as particles are expelled from these charge-starved zones, as we observe in reconnecting current sheets in our simulations (Figure~\ref{fig:rho}).

Charge depletion is seen in all the simulations in Figure~\ref{fig:rho}, independent of the 
resistivity law. To understand the effect, we evaluate the Lagrangian time derivative of $n/B_z$ using a combination
of Faraday's law and the continuity equation.
We start with $\partial_tB_z = -c(\bnabla\times{\bm E}_\perp)\cdot\hat{z}$, with ${\bm E}_\perp \sim -{\bm\beta}_\perp\times B_z\hat z -
\beta_z\hat z\times {\bm B}_\perp + \eta {\bm J}_\perp$.  This gives 
\be
\begin{split}
{1\over c}\partial_t B_z & \sim -{\bm\beta}_\perp\cdot\bnabla_\perp B_z - B_z\bnabla_\perp\cdot{\bm\beta}_\perp
    + {\bm B}_\perp\cdot\bnabla_\perp\beta_z \\
    & \quad -\hat z\cdot [\bnabla\times(\eta {\bm J}_\perp)].
\end{split}
\ee
Since the system is invariant in $z$, the continuity equation takes the form ${1\over c}\partial_tn = 
-{\bm\beta}_\perp\cdot\bnabla_\perp n - n\bnabla_\perp\cdot{\bm\beta}_\perp$.
We confirm that $n/B_z$ is conserved as $v_z$, $\eta \rightarrow 0$, and that to leading order
\begin{align}
    {d\over dt}\left({n\over B_z}\right) \sim -c{n\over B_z^2}
    \left[{\bm B}_\perp\cdot\bnabla_\perp\beta_z + 
    \bnabla_\perp\cdot\left({\eta c\over 4\pi}\bnabla_\perp B_z\right)\right]. \label{eq:dt_n_Bz}
\end{align}

Let us consider first the case of uniform resistivity $\eta$.  Then the two terms on the 
right-hand side of Equation~\eqref{eq:dt_n_Bz} have opposing signs. We find that $B_z$ attains a local maximum in the current layer and thus the second term acts to increase density in the layer.  We find that the first term, which depends on $\beta_z$, dominates and is negative and thus depletes density in the layer. 
This term acts within the boundaries of the outflows from
X-points that form within reconnecting current sheets.  The vertical motion $\beta_z$
is concentrated in the charged vortex sheets bounding such an outflow (moving with velocity
${\bm\beta}_{\rm out} c$, see Figures~\ref{fig:khi} and \ref{fig:plasmoid}) and is driven by the vertical component of the electrostatic force $\beta_z \propto q E_z$ (see Figures~\ref{fig:qEz} and \ref{fig:vz}),
\be
\begin{split}
q E_z & \sim \left( {\bnabla_\perp\cdot {\bm E}_\perp\over 4\pi} \right)\eta J_z \\
&= - {c \eta\over(4\pi)^2}\bnabla_\perp\cdot({\bm\beta}_{\rm out}\times B_z\hat z) 
(\bnabla_\perp\times{\bm B}_\perp)\cdot\hat z.
\end{split}
\ee
The sign of the quantity $-{\bm B}_\perp\cdot\bnabla_\perp\beta_z$ therefore is independent of
the sign of ${\bm B}_\perp$, and as the outflow diverges away from the X-point we find
that this sign is negative:  ${\bm E}\times{\bm B}$ drift within the outflow
causes $n/B_z$ to decrease further.  The second term on the right-hand side of Equation~\eqref{eq:dt_n_Bz} has
the opposing sign, but is subdominant as it depends on the smoother profile of 
$B_z$.

Now let us consider the case, relevant to simulation \texttt{Rev2DNonUni}, where 
$\eta$ varies greatly across the sheet,
$|\bnabla_\perp \eta|/\eta \gg |\bnabla_\perp^2 B_z|/|\bnabla_\perp B_z|$.
This condition may be rewritten approximately as $\ln(\eta_{\max}/\eta_0) \gg 1$, where $\eta_{\max}$ is the peak in the sheet and we have assumed both quantities to vary over the current sheet width. We observe $\eta_{\max} \sim (10^2-10^3) \eta_0$ (see
Appendix~\ref{app:non-uni})\footnote{Importantly, the value of $\eta_{\max}$ does not depend on the choice of $\eta_0$ \eqref{eq:etamod}. Therefore, the ratio $\ln(\eta_{\max}/\eta_0)$ serves as an indicator of an asymptotically small $\eta_0$.}, which corresponds to $\ln(\eta_{\max}/\eta_0)  \sim 5-7$, thus supporting this approximation. 
In such a case, Equation~\eqref{eq:dt_n_Bz} reduces to
\begin{equation}
\begin{split}
    {d\over dt}\ln\left( \frac{n/B_z}{n_0/B_g}\right) &\simeq -
    {c\over B_z}{\bm B}_\perp\cdot\bnabla_\perp\beta_z \\
    & \quad -\bnabla_\perp\left({\eta c^2\over 4\pi}\right)\cdot\bnabla_\perp \ln \left( \frac{B_z}{B_g} \right),
\end{split}
\end{equation}
and the second term also contributes to the charge depleted if
$\bnabla_\perp \eta \cdot \bnabla_\perp B_z > 0$.
In our simulations, $\eta$ and $B_z$ attain a maximum together across the
charge-starved current sheets (\citealt{Ripperda2026}; see
Appendix~\ref{app:non-uni} for plots of $\eta$).

\section{Summary and Application to Magnetars}\label{sec:discussion}

This paper has investigated magnetic reconnection on small scales in the magnetosphere of a neutron star. The non-potential magnetic field is weak compared with the background field but may sustain a significant net twist, remaining for a limited time in a metastable state. This process may underlie the extreme variability that is seen in radio-pulsing magnetars and some bright magnetar X-ray bursts.

We study the response of a smoothly twisted, high-magnetization flux structure to a strong, localized perturbation. Impulsive reconnection is ignited in a situation where spontaneous tearing is slow. The reconnection process is well represented by a planar calculation when the flux structure is very narrow compared with the stellar radius.  Our explorations are based on relativistic, resistive MHD simulations with the largest computationally feasible dynamical range ($\sim$ four-orders of magnitude of resolved scales); this 2D approach is validated using lower-resolution 3D simulations (Appendix~\ref{app:3D}).

Two types of 2D simulation are presented: one with a standard uniform resistivity and Lundquist number $\sim 10^5$; and two simulations that resolve small-scale zones of charge starvation using a recently developed resistivity prescription \citep{2025ApJ...978L..45M, Ripperda2026}. This dual approach is motivated by the wide range of kinetic (charge starvation) scales that are expected outside a magnetar, depending on whether it is strongly bursting or in a more quiescent dissipative state.

We summarize several qualitative features of the reconnection process:
\begin{enumerate}[label=(\roman*)]
    \item The onset of reconnection is sensitive to the placement of the perturbation. The process considered therefore may have a stochastic nature depending on the source of the perturbation (e.g., a localized yielding event in a magnetar crust, or a localized perturbation driving a `dwarf' pulse in a radio pulsar).  The dwarf pulses in particular are a distinct short-timescale and low-energy phenomenon appearing in aging pulsars which frequently enter a radio-quiet nulling state \citep{Yan2024}; only 
    in a fraction of pulses does the small scale event end up triggering a broader radio pulse.
    \item As found previously in 2D and 3D simulations, reconnection cancels off a subdominant sign of twist without mediating a full relaxation to a state of uniform $J_z/B_z$ (a completely Taylor-relaxed state) \citep{Pontin2020}. We find $J_z \simeq j_z$ to vary both in sign and greatly in magnitude at late times, with a nearly uniform $B_z$ (Figure~\ref{fig:jz_panel}).
    \item Reconnection is faster when the effects of small-scale charge starvation are taken into account, consistent with previous representations of reconnection near pulsars and black holes (\citealt{2023ApJ...943..105H,Bransgrove:2021,Ripperda2026}; Figure~\ref{fig:rec_rate}).
    \item Strong intermittency is observed in the resistive heating that is triggered by reconnection. A dominant current sheet forms between two flux elements with opposing twists. A small fraction of the dissipation is concentrated around X-points forming in the current sheet.  Most of the dissipation is driven by KHI of narrow outflows from the current sheet, which become turbulent when colliding with previously ejected fluid (Figure~\ref{fig:khi} and Section~\ref{sec:dynamics}).
    \item The high-wavenumber magnetic spectrum is consistent with $dB_\perp^2 \sim k_\perp^{-2}dk_\perp$, i.e., with a current density that diverges as $\sim k_\perp^{1/2}$ (Figure~\ref{fig:2D_spec}). The origin of this power law, which is tied to the strongly inhomogeneous flow structure, needs further exploration. The role of hydromagnetic instability is highlighted by the fact that a similar power spectrum emerges in lower-resolution simulations  that do not resolve the current sheet and in which the plasmoid instability is absent.
    \item The charge-starved current sheet is partially evacuated by a process that we explain in terms of ${\bm E}\times{\bm B}$ drift (Section~\ref{sec:rho_in_charge_starved_zones}).  The rate of heating per particle in charge-starved zones will be enhanced by this process. 
    \item Peaks in the current density and non-ideal electric field may either coincide with X-points or with KH-unstable zones (Figures~\ref{fig:jz_panel}, \ref{fig:jz_loc}, and \ref{fig:qEz}). In a neutron star magnetosphere with realistic magnetization, these heated zones represent relativistic particle flows along the guide magnetic field. Electromagnetic emission from these localized flows would be strongly modulated by rotation of the neutron star.
\end{enumerate}

Now we discuss some potential applications and implications of this reconnection process for the astrophysics of
magnetars, in particular for radio variability and crustal heating during bright X-ray outbursts.

\subsection{Radio Variability and Charge Starvation in the Outer Magnetosphere}

The transient magnetar XTE J1810$-$197 \citep{Ibrahim2004}, with a $5.54~{\rm s}$ spin period, was the first magnetar to show detectable
radio pulsations \citep{Camilo2006} and has been extensively studied in the radio band.  In particular, individual radio pulses show strong intermittency \citep{Dai2019} and the integrated pulse profile experiences dramatic changes in shape over hundreds of rotations (e.g. \citealt{Huang2023}).  At the same time, the radio polarization exhibits the S-curve swing that is typical of many pulsars \citep{Lower2021}. The radio pulses of XTE J1810$-$197 therefore appear to originate inside its light cylinder, even while the current flowing near the magnetic separatrix is non-axisymmetric and rapidly varying.  

We consider the possibility that magnetic reconnection on the kinetic scale, as examined here through our sub-grid resistivity model, is the source of the strong intermittency within radio pulses.  Figure~\ref{fig:current} shows a series of slices through the current profile in 
simulation \texttt{Rev2D}, representing a rapid transit of the line of sight of an observer through the plasma due to the large-scale
rotational motion of the NS.  Each line of sight crosses zones of high current density only within narrow intervals representing the widths of 
reconnecting current sheets.  Similarly, in
model \texttt{Rev2DNonUni}, zones with enhanced resistivity and strong
implied charge starvation ($\tilde\eta > 0.1$) fill only $\sim 1\%$ of
the volume at the peak of reconnection (Figure~\ref{fig:threshold_areas}).  
In the regime of extreme magnetization expected outside the magnetar,
the microscopic particle momentum distribution must be
nearly aligned with ${\bm B}$, suggesting that electromagnetic emission by particles in current spikes will be beamed toward the observer.

Radio magnetars have been found to share with ordinary pulsars a correlation between pulse microstructure timescale $\Delta t$ and rotation period $P = 2\pi/\Omega$ \citep{Kramer2024}.  This correlation, $\Delta t\sim 10^{-3} P$, has a simple kinematic explanation \citep{T22}: the microstructure is representative of the emitting Lorentz cone of the secondary pairs on the open field bundle (Lorentz factor $\gamma_\pm \sim 50-100$) that are produced in the polar cap of a radio pulsar. As long as the emitting particle beam is sufficiently narrow, the angular width of its emission is $\delta\theta_\pm \sim 1/\gamma_\pm$, which when swept past the line of sight by the stellar rotation gives a temporal width $\Delta t \sim \delta\theta_\pm / \Omega = P/2\pi\gamma_\pm$.  

We require that the geometrical width of the particle beam be narrower than $r_{\rm em}\delta\theta_\pm$, where $r_{\rm em}$ is the emission radius. In this circumstance, the expected width $x_s$ of charge-starved current layers is sensitive to  the particle multiplicity. Near the light cylinder we may take a shear length $L \sim c/\Omega$ and $B_\phi \sim B_g$, in which case the analysis of Section~\ref{sec:starve} shows that $\Omega x_s / c \sim 1/{\cal M}_\pm$. The $e^\pm$ multiplicity is modest (${\cal M}_\pm \ \sim 10-100$) when the pairs are supplied by non-local collisions between MeV energy gamma rays \citep{Zhang2025}, and somewhat higher ($\sim 10^2-10^3$) when the main particle source is resonant cyclotron scattering of keV blackbody photons within several stellar radii of the magnetar surface \citep{B13}. Taking the emission to occur at a radius $r_{\rm em} \sim 0.1c/\Omega$, as is frequently inferred for ordinary radio pulsars \citep{PK2022}, we have $x_s / \delta\theta_\pm r_{\rm em} \sim (\gamma_\pm/{\cal M}_\pm) (\Omega r_{\rm em}/c)^{1/2} \sim 0.3 (\gamma_{\pm,2}/{\cal M}_{\pm,2})  < 1$.  This ratio is yet smaller if we substitute a higher multiplicity ${\cal M}_\pm \sim 10^{4-5}$ that is inferred for young active pulsars \citep{Timohkin2019}. We conclude that the width of the observed radio microstructure is consistent with emission from a large collection of kinetic-scale emitting elements, similar to those realized in Figure~\ref{fig:current}.

\subsection{Crustal Heating and Matter Expulsion}\label{sec:surface}

We have described how magnetic reconnection induces volumetric heating within a slender, relativistic flux tube. The magnetosphere of a bursting magnetar may be described as a collection of such flux tubes with variable signs of twist. Some fraction of the magnetic energy dissipated is initially trapped in optically thick electron-positron-photon plasma, as seen in the pulsating tails of the giant flares \citep{Feroci01}.  

The mechanism of current damping described here extends below the surface of the magnetar, at least to a mass column $\Sigma$ where the hydrostatic pressure $\Sigma g$ in gravitational field $g$ is comparable to the thermal pressure $p_{\rm diss}$ released by the dissipated magnetic field.  During a giant outburst, the plasma trapped inside the magnetosphere contains enough energy to reach a temperature $T \sim 1$ MeV, thereby compressing the outer crust \citep{TD95}. The total electron density and kinetic scale $d_e$ are therefore comparable in the heated magnetosphere and the compressed outer crust;  the damping of small-scale currents by reconnection should proceed similarly in these two zones.

Two interesting observational diagnostics are suggested by this process of direct, volumetric heating and melting of the outer crust. First, after the external fireball dissipates, the heated crustal layer re-expands and re-radiates the trapped energy over a much longer timescale, in the form of softer surface X-ray emission and deeper neutrino cooling \citep{2002ApJ...580L..69L}. Second, deep heating of the magnetar crust opens up the possibility of ejection of baryonic material during an extended outburst, such as a giant magnetar flare.  Evidence for the ejection of $10^{25-26}$ g of material in the 2004 flare of SGR 1806$-$20 (the most energetic Galactic magnetar flare yet detected) comes from a weeks long radio synchrotron afterglow \citep{Granot2006}.

In response to multiple heating events, we now explain how an outer layer of the crust will be levitated sufficiently to escape from the star when the magnetic field opens up. Representing the twisted field by a cylinder of radius $a$ and length $L_z$, the toroidal field reaches $B_\phi \sim \pi (a/L_z) B_z$.   The characteristic surface mass column that is strongly heated is then
\begin{equation}
    \begin{split}
    &\Sigma_{\rm eq} = \frac{1}{g} p_{\rm diss} \\
    &= 2.5\times 10^{14}\, g_{14}^{-1} 
       \left(\frac{a}{0.1L_z}\right)^2
       \left(\frac{B_z}{100\,B_{\rm Q}}\right)^2
       \, {\rm g~cm^{-2}},\label{eq:sigheat}
    \end{split}
\end{equation}
where $B_{\rm{Q}} = 4.4 \times 10^{13}$~G is the Schwinger field and we have taken the fireball pressure $p_{\rm diss}$ to be $1/3$ of the dissipated toroidal
magnetic energy density.  The corresponding mass density and depth at the base of the heated layer 
are $\rho \sim 10^{10}$ g cm$^{-3}$ and $|z| \sim 200$ m.  This means that charge starvation effects
only become important on scales vastly smaller than the depth of the heated layer, by more than 10 orders of magnitude.

The ion mass dominates the gravitating mass even when the electrons are heated sufficiently to become non-degenerate, thereby allowing a comparable density of positrons to form deep within the heated surface layer.  At the peak of fireball pressure, the matter pressure is therefore increased by a factor $1+\varsigma^{-1}$, where $\varsigma\equiv \Sigma/\Sigma_{\rm eq}$.  After the fireball pressure dissipates, the layer expands until its pressure returns to $\Sigma g$. In the outer layers of the neutron star crust, density scales with pressure as $\rho \propto p^{3/4}$. The reduction in ion density $\rho$ after one heating episode is $\rho_1/\rho_0  =(1+\varsigma^{-1})^{-3/4}$.  The net increase in thickness after $n$ heating episodes is $z_n / z_0 = \rho_0 / \rho_n = (1+\varsigma^{-1})^{3n/4}$. The column \eqref{eq:sigheat} corresponds to an initial depth $z_0 \sim 200$ m, and so expansion to a height $z$ comparable to the NS radius $R_{\rm NS}$, followed by escape from the magnetar, requires $n \sim 10$ heating episodes for $\varsigma \sim 1$.

The magnetar will eject as much as $\sim 10^{26}$ g if this process covers $\sim 10\%$ of the magnetar surface. Such a large covering fraction appears plausible in the 2004 giant flare, which released more than $10^{46}$ erg over an interval of $t_{\rm spike} \sim 0.2$ s \citep{Tanaka2007}. Given that the crust is displaced horizontally with a speed $v_{\rm frac}$ comparable to the shear wave speed, $v_{\rm frac} = \varepsilon_\mu v_\mu \sim \varepsilon_\mu 10^8$ cm s$^{-1}$, the displacement along the magnetar surface in the 2004 event was at least $200\,\varepsilon_\mu$ km. 

The requirement that each part of the heated magnetosphere experiences at least $\sim 10$ opening/closing events then points to the presence of higher magnetic multipoles of dimension a few km. Such small-scale magnetic structure would allow a displacement of $\sim 30$~km, corresponding to roughly half a stellar rotation, to produce $\sim 10$ opening/closing events. Direct evidence for the presence of multipoles in giant flare sources comes from the complicated structure of the X-ray pulse profile during and following outbursts, and indirect evidence from the large cumulative energy release in flares over their active lifetimes \citep{Feroci01,TD01}.  %Ab initio calculations of magnetic relaxation in neutron stars show that such structures form naturally \citep{Li2026}.  

This ejection process is a direct consequence of the shearing mechanism that appears to underlie much of magnetar 
activity and removes the need to invoke global magnetic buoyant escape from a thin outer layer of the crust as a driver of the most energetic flares \citep{Bransgrove2026}, or the formation of downward propagating shock waves depending on large-scale rearrangements of the magnetosphere
on microsecond timescales \citep{Cehula2024}.

\section*{Acknowledgments}

The authors thank Andrei M. Beloborodov, Alexander Philippov, Sean Ressler, James M. Stone, and Jonathan Zhang for insightful discussions. 

LLM usage disclosure: Anthropic's Opus 5.5 and OpenAI’s GPT-6 were used to assist with manuscript editing and with the development of post-processing and visualization code. All LLM generated text and code were verified by the authors. The scientific content, methods, and conclusions were developed and verified by the authors, who take full responsibility for the final manuscript.

This research was supported in part by grant NSF PHY-2309135 to the Kavli Institute for Theoretical Physics (KITP). 

We acknowledge the support of the Natural Sciences and Engineering Research Council of Canada (NSERC), CGS D-588952-2024 (M.~P.~G.), funding reference numbers RGPIN-2023-04844 and ALLRP 592744 (B.~R.), and funding reference number RGPIN-2023-04612 (C.~T.). C.T. and B.R.~are supported by a grant from the Simons Foundation (MP-SCMPS-00001470). B.R. is supported by the Canadian Space Agency (23JWGO2A01). The computational resources and services used in this work were partially provided by facilities supported by the VSC (Flemish Supercomputer Center), funded by the Research Foundation Flanders (FWO) and the Flemish Government – department EWI, and by Compute Ontario and the Digital Research Alliance of Canada (alliancecan.ca) compute allocation rrg-ripperda and rrg-thompsoc. J.~R.~B. also acknowledges support from NSF Award 2206756 and the Natural Sciences and Engineering Research Council of Canada (NSERC), funding reference number 568580. During preparation of this manuscript, J.~R.~B. was supported by NASA through NASA Hubble Fellowship grant HST-HF2-51597.001-A, awarded by the Space Telescope Science Institute and administered by AURA for NASA under contract NAS5-26555.

\appendix

\section{Explicit Field Configurations}\label{app:Configurations}
The vector potential $A_z$ corresponding to  Equation~\eqref{eq:Bphi1} that is used to define the initial condition is
\begin{equation}
    A_z = -{K_{\phi1}\over 4} B_0 R \tr^2 (\tr - 1)^2, \quad (r\leq R)
\end{equation}
with $\tr \equiv r/R$ and $A_z = A_z(R) = 0$ outside the flux tube.  The solution to Equation~\eqref{eq:ff} for $B_z$ is
\begin{equation}
\begin{split}
        B_z(r) = & \left[ B_g^2-\frac{18}{5} B_0^2 \tr^2 \left(585 \tr^2 - 504 \tr^3 \vphantom{\frac{1}{2}}+160 \tr^4-300 \tr+60 \right) \right]^{1/2},\quad (r < R),
\end{split}
\end{equation}
where $B_g$ is the peak value of the guide field at $r=0$, and $B_z = \sqrt{B_g^2 - (18/5) B_0^2}$ outside the flux tube.

\section{Three-Dimensional Simulations}\label{app:3D}

In this section, we present simulations of a 3D flux tube with the same twist profile as the 2D model described in the main text.  Two simulations are performed, one with a guide field of the same relative strength ($B_g = 5 B_0$, model \texttt{Rev3D}) and a second with a weaker guide field ($B_g = 2.25 B_0$, model \texttt{Rev3DKink}).  The evolution with $B_g = 5 B_0$ remains nearly planar, maintaining approximate translational symmetry along the longitudinal $z$-coordinate. After application of a localized perturbation, the flux tube with weaker guide field supports an ideal kink mode that grows faster than the in-plane triggered reconnection seen in 2D, thereby breaking translational symmetry.  

These 3D simulations solve the equations of ideal, relativistic MHD using the code \texttt{AthenaK} \citep{2024arXiv240916053S, 2025ApJS..278...50Z, 2025ApJS..276...35F}. They employ zero gradient boundary conditions in the $z$-direction, so as to simulate an isolated section of a much longer flux tube, which is causally disconnected from either end on the reconnection timescale. Simulations are initiated with uniform density, uniform dimensionless temperature $T_0 = p_0/\rho_0 c^2 = 1$, and magnetization $\sigma =10$ at $r=0$.  The effective resolution is $1024^2\times 2560$ over a domain comprising $4.8R\times 4.8R$ in the plane and vertical extent $L_z/2R = 5$. The size of the in-plane domain was increased in 3D in order to accommodate the use of static mesh refinement and the larger mesh blocks  needed to utilize GPUs effectively. The magnetization was reduced by a factor 10 compared with the 2D simulations to compensate the reduction in numerical stability we find in the lower resolution 3D simulations.

\begin{figure*}
    \centering
    \leavevmode
    \begin{tikzpicture}
    % Lower image
    \node[inner sep=0] (lower) at (0,-0.455\linewidth) {
        \includegraphics[width=\linewidth]{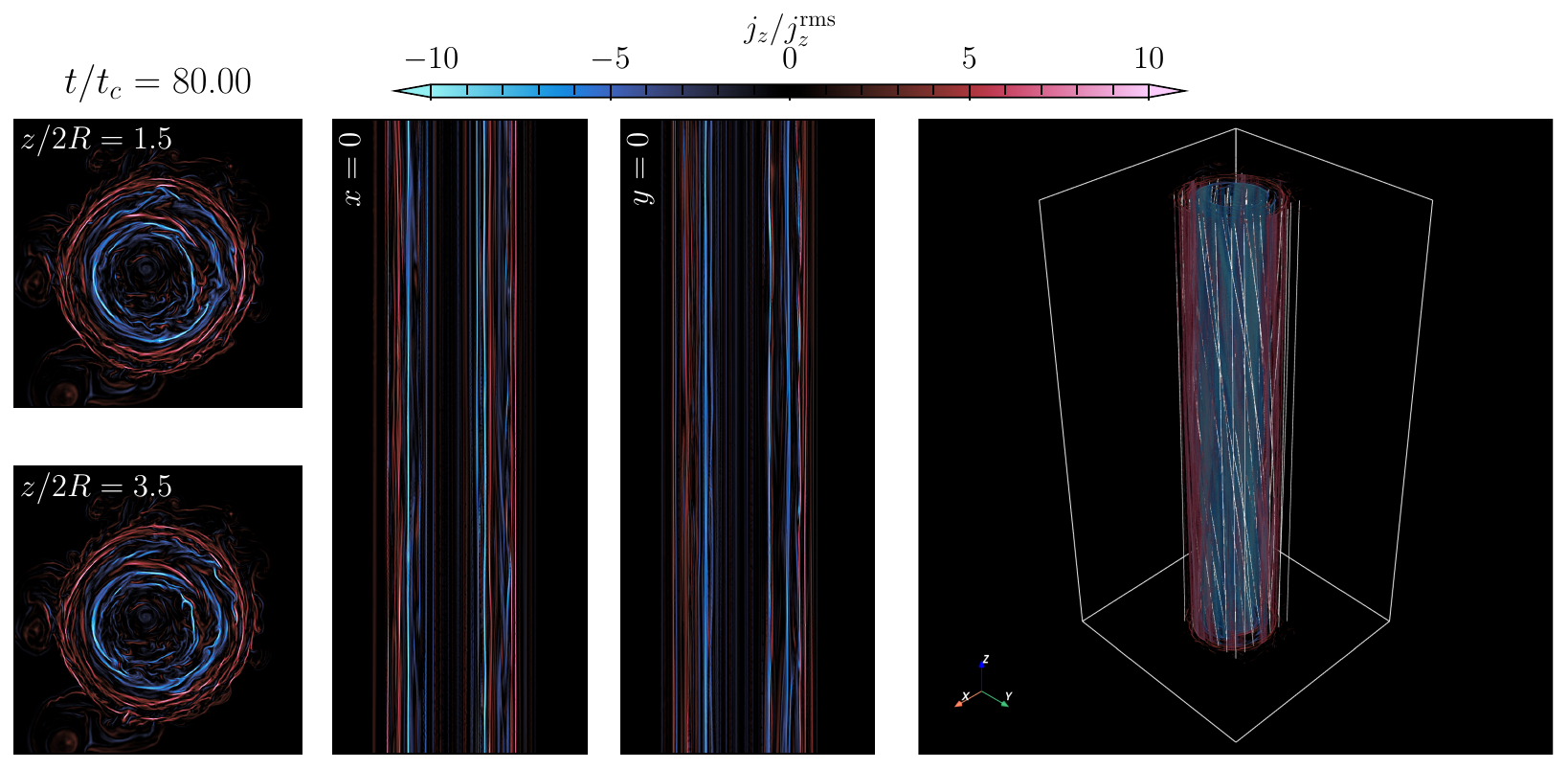}
    };
    % Upper image -- drawn LAST = higher z-order
    \node[inner sep=0] (upper) at (0,0) {
        \includegraphics[width=\linewidth]{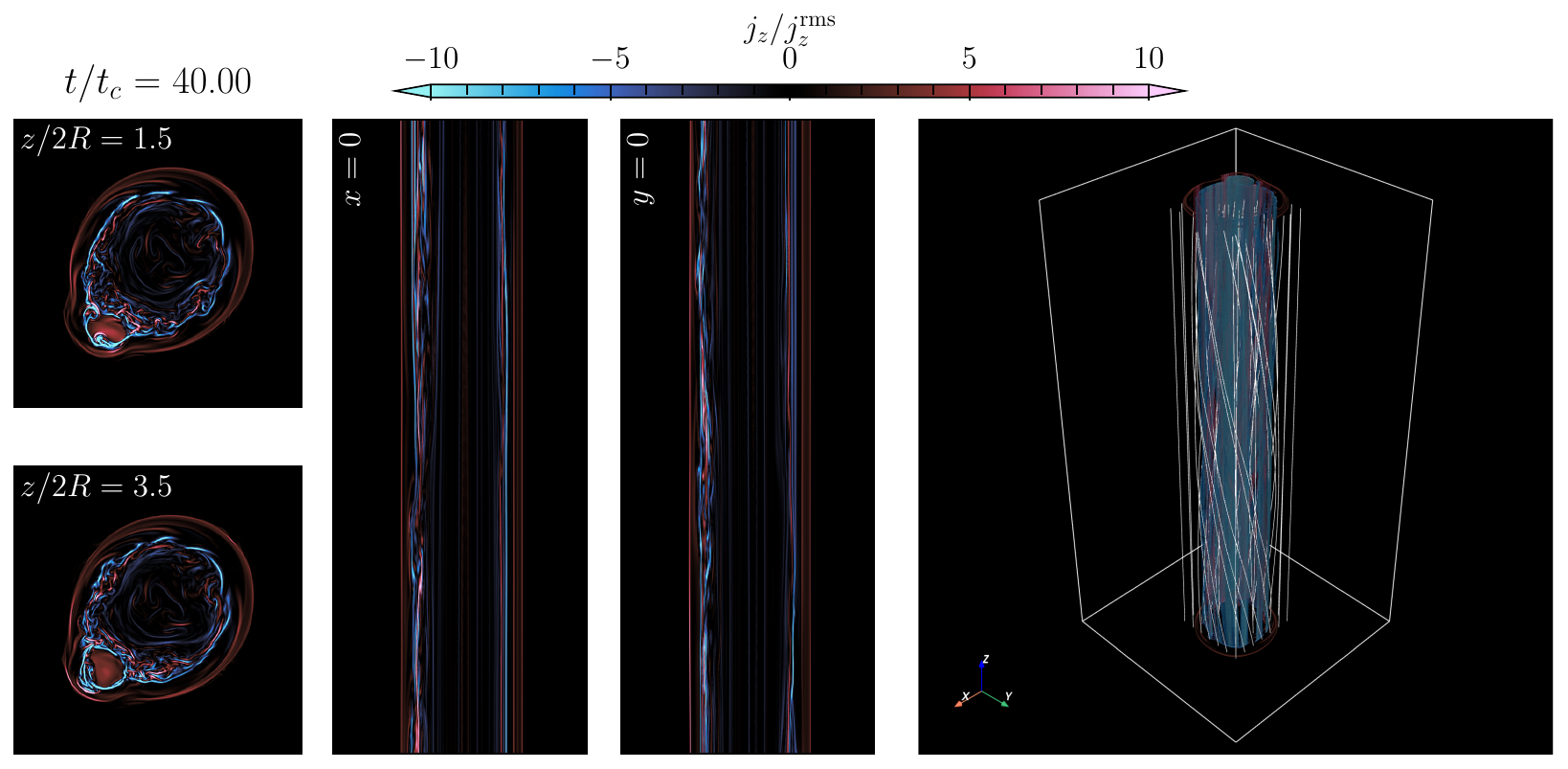}
    };
    % Mask
    \fill[white]
        (lower.north east)
        ++(-15cm,-0.6cm)
        rectangle
        ++(15cm,-0.75cm);
    \end{tikzpicture}
    \caption{
    Two snapshots of $j_z = (c/4\pi)(\bnabla\times\bm{B})_z$, in 3D simulation \texttt{Rev3D}, normalized by the r.m.s. value in the computational domain. The first column shows cuts in the $xy$-plane at the indicated value of $z$. The second and third column are cuts through $x=0$ and $y=0$. The fourth column is a 3D rendering with opacity scaling with the magnitude of the current. Magnetic field lines are drawn in white. The strong guide field suppresses the growth of structure in the $z$-direction and the dynamics are dominated by reconnection in the $xy$-plane.}
        \label{fig:3D_nokink}
\end{figure*}

\begin{figure*}
    \centering
    \leavevmode
    \begin{tikzpicture}
    % Lower image
    \node[inner sep=0] (lower) at (0,-0.455\linewidth) {
        \includegraphics[width=\linewidth]{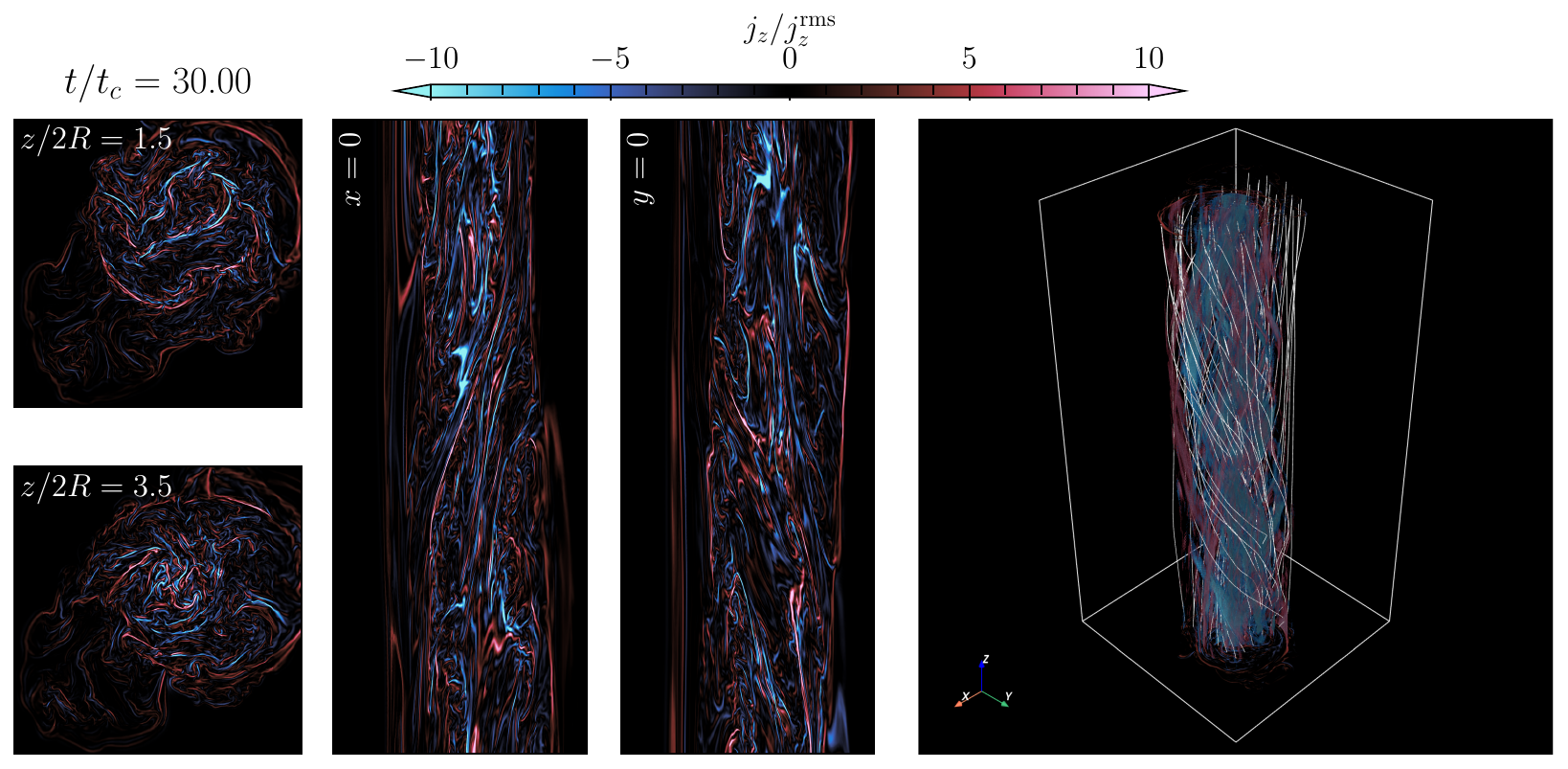}
    };
    % Upper image -- drawn LAST = higher z-order
    \node[inner sep=0] (upper) at (0,0) {
        \includegraphics[width=\linewidth]{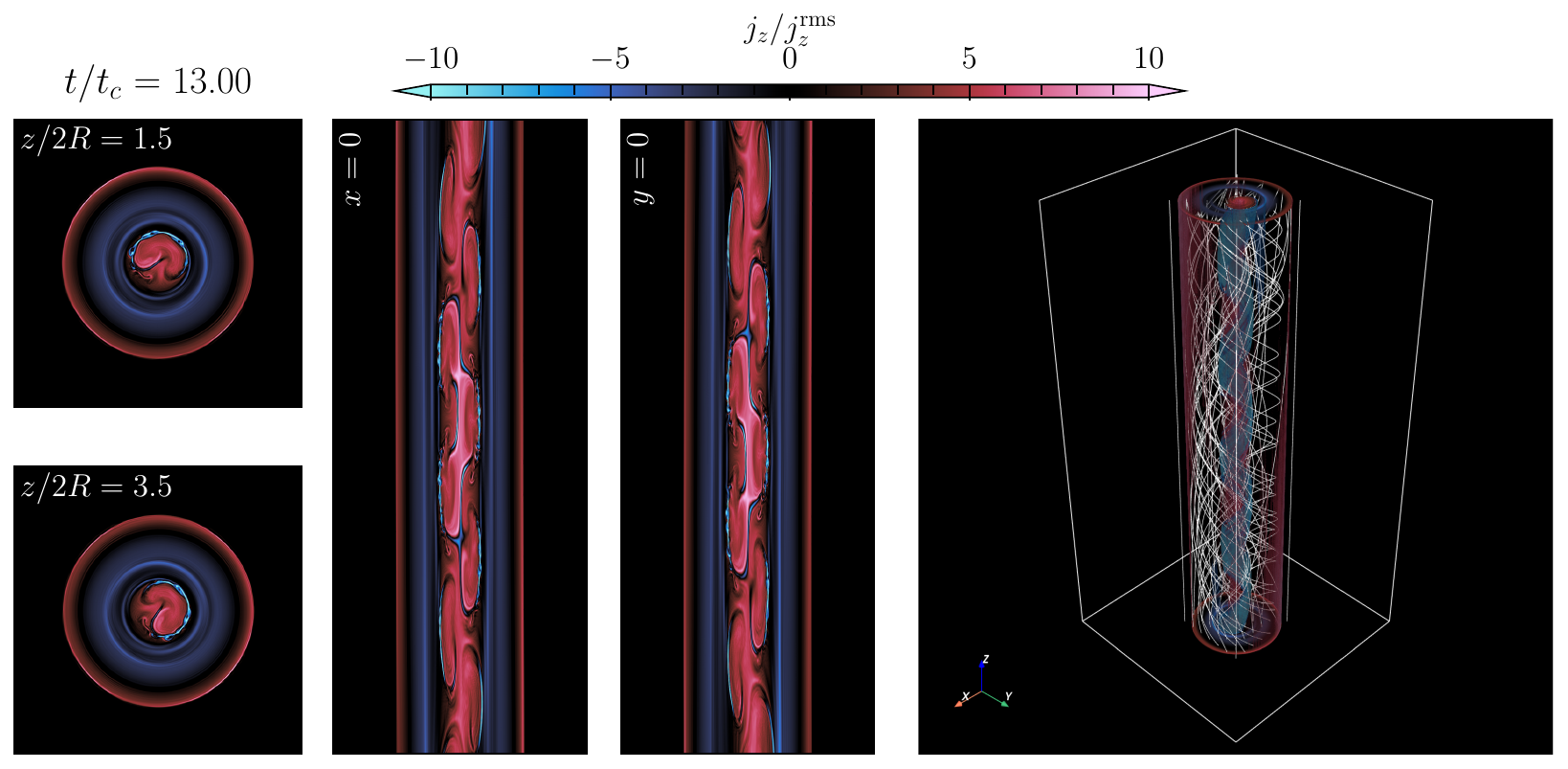}
    };
    % Mask
    \fill[white]
        (lower.north east)
        ++(-15cm,-0.6cm)
        rectangle
        ++(15cm,-0.75cm);
    \end{tikzpicture}
    \caption{
    Same as Figure~\ref{fig:3D_nokink} but for 3D simulation \texttt{Rev3DKink}. The weaker guide field results in a faster kink mode and the dynamics become a mix of the growing kink mode and reconnection which is not restricted to the $xy$-plane. Significant structure grows in the $z$-direction.
    }
    \label{fig:kink}
\end{figure*}

Consider first model \texttt{Rev3D} with $B_g/B_0 = 5$, which has $Q$ below unity only near $r=0$, $L_z/2R=B_g/B_0$. As shown in Figure~\ref{fig:3D_nokink}, reconnection is triggered by the localized perturbation uniformly in $z$, while very little structure develops in the $z$-direction.  This is as expected, given that the $m=1$ ideal kink mode has a growth rate $\Gamma_{\rm kink} \sim 0.6 t_{\rm rec}^{-1}$ \eqref{eq:rec_vs_kink_rate}. A reconnecting current layer has emerged at time $t/t_c = 40$ over a range of coordinate $z$. In the later snapshot, $t/t_c=80$, reconnection is nearly complete and almost all of the interior twist has dissipated.  The current profile has settled into a series of nested circular sheets which continue slowly to diffuse.  This evolution closely resembles that seen in the 2D resistive MHD simulation \texttt{Rev2D} (see the first row of Figure~\ref{fig:jz_panel}).  This resemblance between ideal and resistive MHD representations of reconnection is expected based on previous calibration experiments \citep{2025arXiv250320013G}.

Strong growth of the $m=1$ kink mode is observed in model \texttt{Rev3DKink} with a weaker guide, $B_g/B_0=2.25$ (Figure~\ref{fig:kink}). Now the growth of the kink as predicted by Equation~\eqref{eq:rec_vs_kink_rate} is relatively fast, $\Gamma_{\rm kink} \sim 3t_{\rm rec}^{-1}$. The evolution following the imposition of the perturbation is fully three-dimensional: the inner flux rope is periodically displaced to the side, and the system shows a combination of kink growth and reconnection. Forming current sheets develop curvature along the $z$-axis at early times (Figure~\ref{fig:kink}). At late times, $t/t_c = 30$, the system settles into a configuration of complicated small scale current sheets with structure in the vertical direction. 

We conclude that, when reconnection is ignited by a local perturbation, the 3D nature of the flux tube becomes relevant if the guide field is not too strong, so that $\Gamma_{\rm kink}t_{\rm rec} \gtrsim 1$. Flux structures of a small width must violate this condition, thereby motivating our high-resolution 2D experiments. The evolution emerging from the strong guide-field simulation \texttt{Rev3D} is nearly identical to the corresponding 2D simulations, thereby justifying our focus on 2D simulations that can resolve a larger scale separation.

\begin{figure*}
    \centering
    \includegraphics[width=1.0\linewidth]{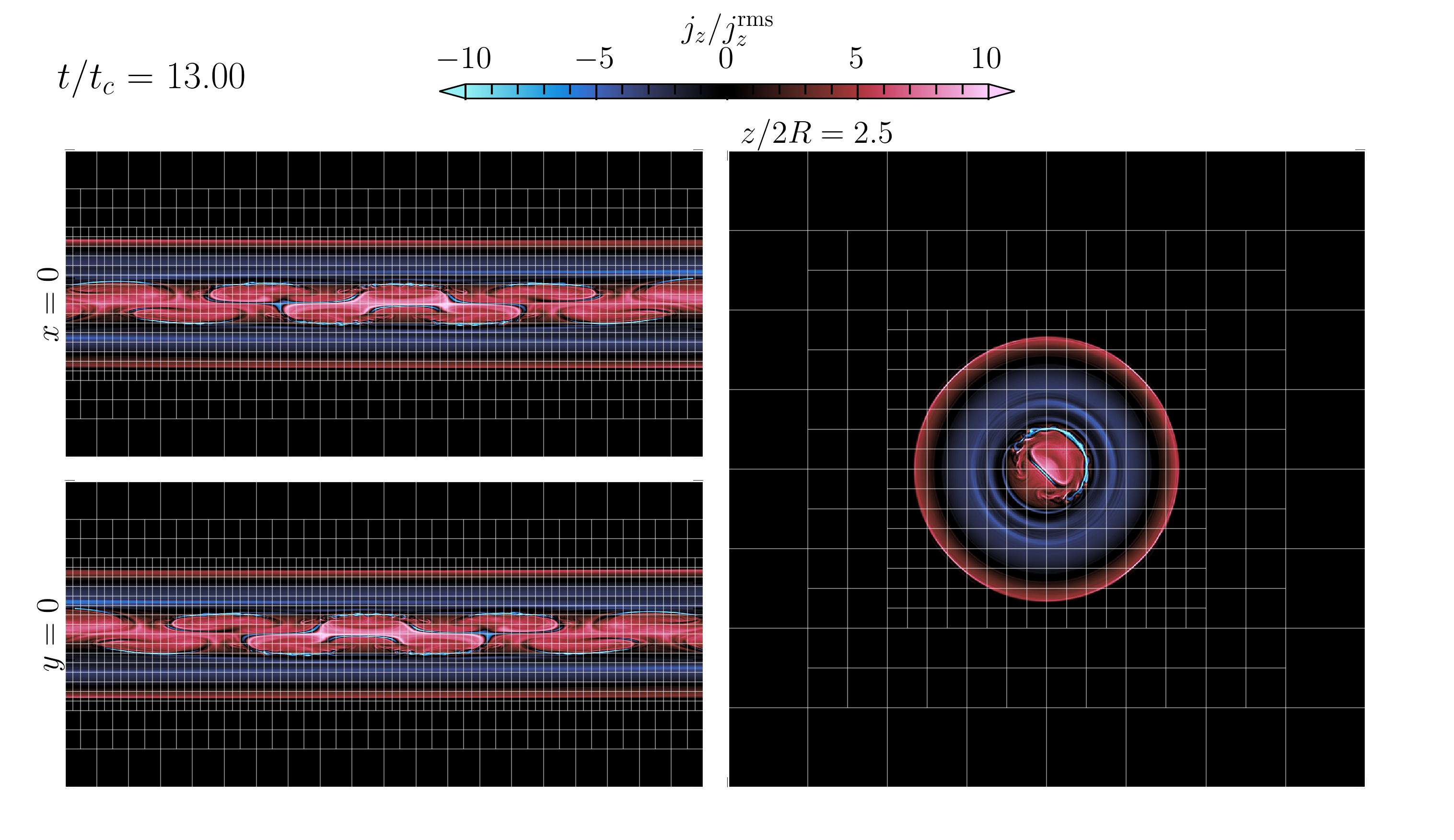}
    \caption{A single snapshot of \texttt{Rev3DKink} showing $j_z = (c/4\pi)(\bnabla\times\bm{B})_z$ and the 3 levels of static mesh refinement. The mesh refinement is concentrated within $|x|,\, |y| \leq 11 R /10$. The first column shows cuts through $x=0$ and $y=0$, while the second column is the $xy$-plane cut through the middle of the $z$-extent $z/2R = 2.5$. The current density is normalized by the r.m.s. value in the plotted domain. Mesh blocks of $32\times 32 \times 32$ cells are outlined in white.}
    \label{fig:amr_3d}
\end{figure*}

\begin{figure*}
    \centering
    \includegraphics[width=1.0\linewidth]{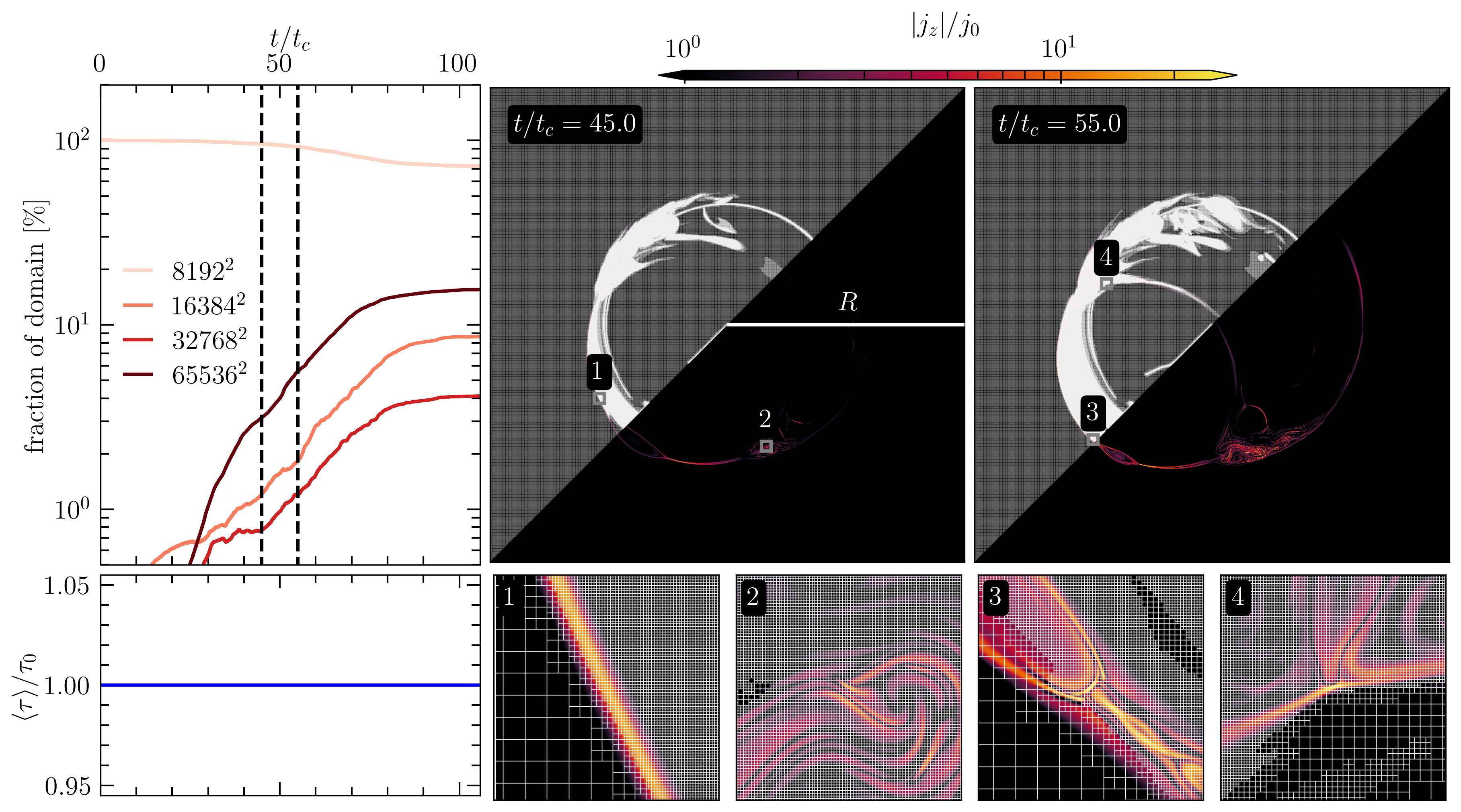}
    \cprotect
    \caption{
    Top left: the area of the domain $r\leq\sqrt{2} R$ occupied by cells of different mesh refinement levels, labeled by the effective resolution over the full computational domain. Vertical lines are plotted to indicate the times plotted in the 2D color plots. Bottom left: the domain averaged conserved rescaled energy density, $\tau = U-D$, where $U$ is the energy density and $D=\rho \Gamma$ is the lab frame mass density, normalized by the initial value $\tau_0 = \langle \tau (t=0) \rangle$. Middle and right: visualizations of $j_z = (c/4\pi) (\bm{\nabla}\times\bm{B})_z$ normalized by the equilibrium maximum, plotted for two times. Gray rectangles indicate the zoomed in regions plotted in the bottom row, which are labeled with corresponding numbers. Mesh blocks of $8\times 8$ cells are outlined in white. Data are from simulation \Verb+Rev2D+.}
    \label{fig:amr}
\end{figure*}

\section{Adaptive and Static Mesh Refinement}\label{app:amr_res}

To minimize computational expense, while resolving the small scale dynamics of the 2D simulations, we use a modified version of the AMR scheme presented in \cite{2021ApJ...923L..13C}. We define lengths $\ell_{\bm{B}}$ and $\ell_{\bm{v}}$, associated with the magnetic twist and vorticity,
\begin{equation}\label{eq:length}
\ell_{\bm{B}} = \frac{\sqrt{B_\perp^2}}{|(\bnabla\times\bm{B})_z|} 
    ,\quad
 \ell_{\bm{v}} = \frac{\sqrt{v_\perp^2}}{|(\bnabla\times\bm{v})_z|}.
\end{equation}
To avoid resolving areas with weak currents but also near zero in-plane field, we require that $|(\bnabla\times\bm{B})_z| \geq  10 \, B_0/(2R)$ and $|(\bnabla\times\bm{v})_z| \geq 5 \,c/(2R)$ before the AMR routine is called. The grid is refined whenever either $\ell_{\bm B}$ or $\ell_{\bm v}$ is not resolved by 8 grid cells in either direction.  To accommodate the large domain, which acts to isolate the flux tube from boundary conditions, we enforce static mesh refinement inside a region enclosing the flux tube, $r \leq\sqrt{2} R$, at a level higher than in the surrounding current-free domain. The entire active flux bundle must be well resolved for the AMR scheme to be effective. For this reason, the 3D simulations only make use of static mesh refinement (Figure~\ref{fig:amr_3d}); the AMR scheme is ineffective for resolutions around the maximum resolution of the 3D runs. Figure~\ref{fig:amr} demonstrates how this mesh refinement scheme effectively tracks small scale currents and vorticity.  The use of AMR significantly reduces the computational expense of the higher resolution 2D simulations.  As reconnection begins, and small scale currents are ejected from the dominant current sheet, a growing portion of the computational domain is filled with the highest levels of mesh refinement.  The computation therefore slows down at later times. Figure~\ref{fig:amr} also demonstrates the energy conserving properties of the code. 

\section{Numerical Convergence}\label{app:con}

\begin{figure*}
    \centering
    \includegraphics[width=\linewidth]{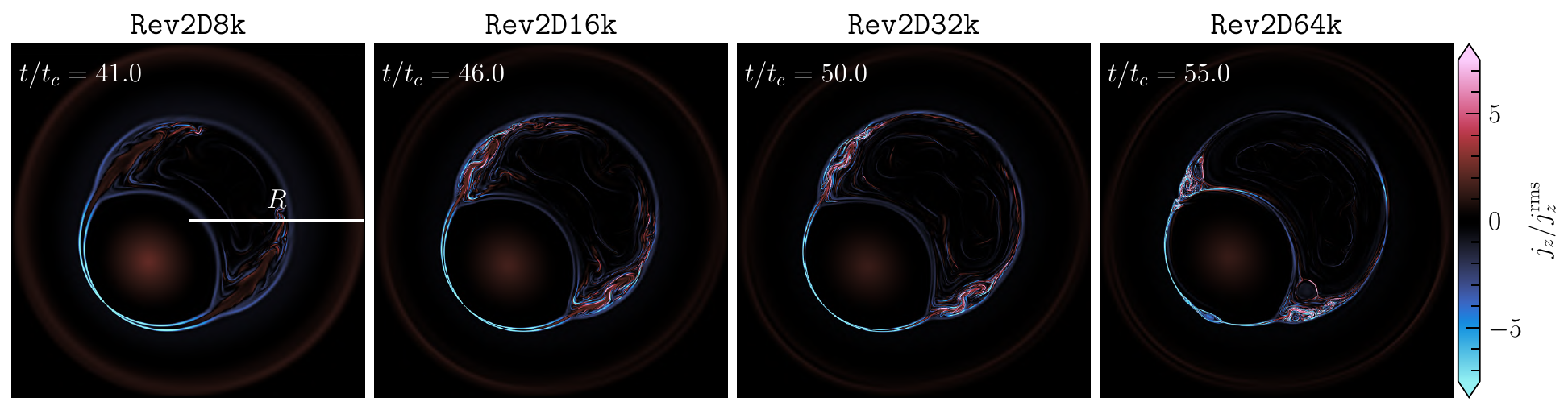}
    \caption{The current density associated with the magnetic curl $j_z = (c/4\pi) (\bnabla\times\bm{B})_z$, normalized by the r.m.s. value in the domain $|x|, |y| \leq R$, for four resolutions ($N_\perp \in \{ 8192, 16384, 32768, 65536\}$) of the same 2D flux tube configuration with uniform resistivity. Time snapshots are chosen such that $\langle B_\perp^2 \rangle/B_0^2$ is nearest to a set value, so that all the simulations are at a similar stage of evolution. The plasmoid instability is suppressed at lower resolution.}
    \label{fig:app1}
\end{figure*}

\begin{figure*}
    \centering
    \includegraphics[width=\linewidth]{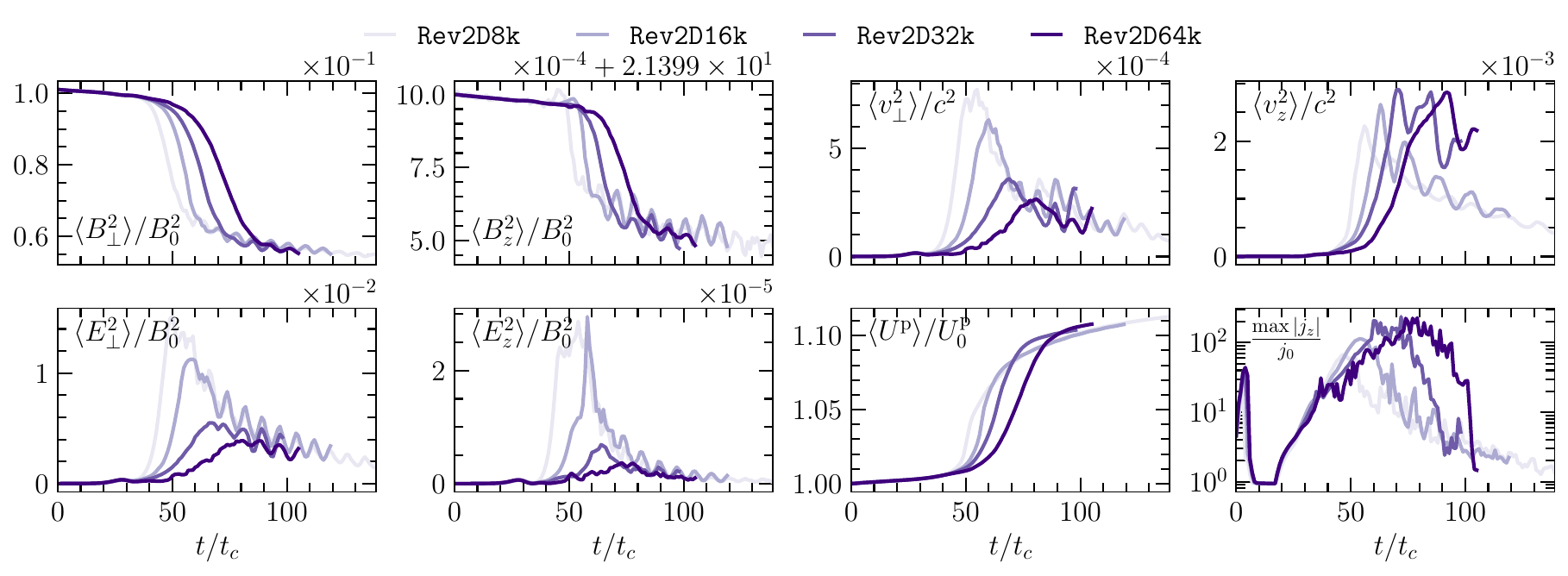}
    \caption{
    Same as Figure~\ref{fig:time_avg} with the addition of the domain averaged in-plane field $\langle B_\perp^2 \rangle$ and out-of-plane field $\langle B_z ^2 \rangle$, now for multiple resolutions, up to the maximum resolution which are the results presented in the main body of the paper.
    All quantities are averaged over the entire computational domain $|x|,|y| \leq 2R$.
    }
    \label{fig:time_avg_con}
\end{figure*}

We have performed a series of uniform-resistivity 2D simulations with the same initial conditions but lower resolution than \texttt{Rev2D}, with $N_\perp \in \{ 8192, 16384, 32768, 65536\}$.   (These are elsewhere denoted as $\{$`8k', `16k', `32k', `64k'$\}$;  note that \texttt{Rev2D}=\texttt{Rev2D64k}.) The plasmoid instability is absent in the three lower-resolution simulations (Figure~\ref{fig:app1}), while the dissipation of the inner twist proceeds in a qualitatively similar manner and at a slightly higher rate. Although reconnection occurs at a singular X-point, the outflow from the current sheet still generates a turbulent flow by exciting a KH instability. This indicates that plasmoid instability is not required to generate a broad spectrum of current perturbations.

The macroscopic dynamics maintains a constant behavior over a wide range of resolution.  One sees this in the time evolution of various domain-averaged quantities, along with the peak current density (Figure~\ref{fig:time_avg_con}). The in-plane magnetic field drops rapidly as the interior sign of twist is dissipated, converting the available magnetic energy into kinetic and thermal plasma energy. The plasma then sloshes in a quasi-periodic manner after the initial burst of reconnection, with these motions remaining contained within the initial current-carrying domain.  Outflows from opposing ends of the current sheet traverse half the circumference before interacting and reversing direction. This results in oscillations with period $\sim 8 t_c$, which matches the crossing time of the circumference at $r=R/2$ (where $B_\phi$ reverses sign initially) at the in-plane Alfv\'{e}n speed, $T \sim \pi (B_g/B_0) (R/c) \sim 7.8 t_c$. The value of the peak current density is a strict measure of convergence in such a non-linear system and sensitive to resolution:  we find $j_{\rm max} \propto 1/\Delta x$, where $\Delta x$ is the grid scale size, up to the two highest resolutions \texttt{Rev2D32k} and \texttt{Rev2D64k} where it appears to saturate. This indicates that the system is marginally resolved in \texttt{Rev2D32k} and well resolved in \texttt{Rev2D64k}, as we argue below this is calculated using the expected sheet thickness.

The half-thickness of the current sheet in a simulation with uniform $\eta$ is approximately $\delta \sim c (\eta/4\pi) (1/\beta_{\rm rec}) (B_g/B_0)$. In the 2D simulation \texttt{Rev2D} of highest resolution, $\delta$ is well resolved by $\sim 10$ grid cells (and only marginally resolved by $\sim 5$ cells in \texttt{Rev2D32k}).  Thus for $N_\perp \lesssim 32768$, the current sheet is under-resolved to marginally resolved and numerical dissipation reduces the effective Lundquist number, thereby suppressing the plasmoid instability. By contrast, the current sheet is well resolved for $N_\perp = 65536$.

\begin{figure*} 
\centering \includegraphics[width=\linewidth]{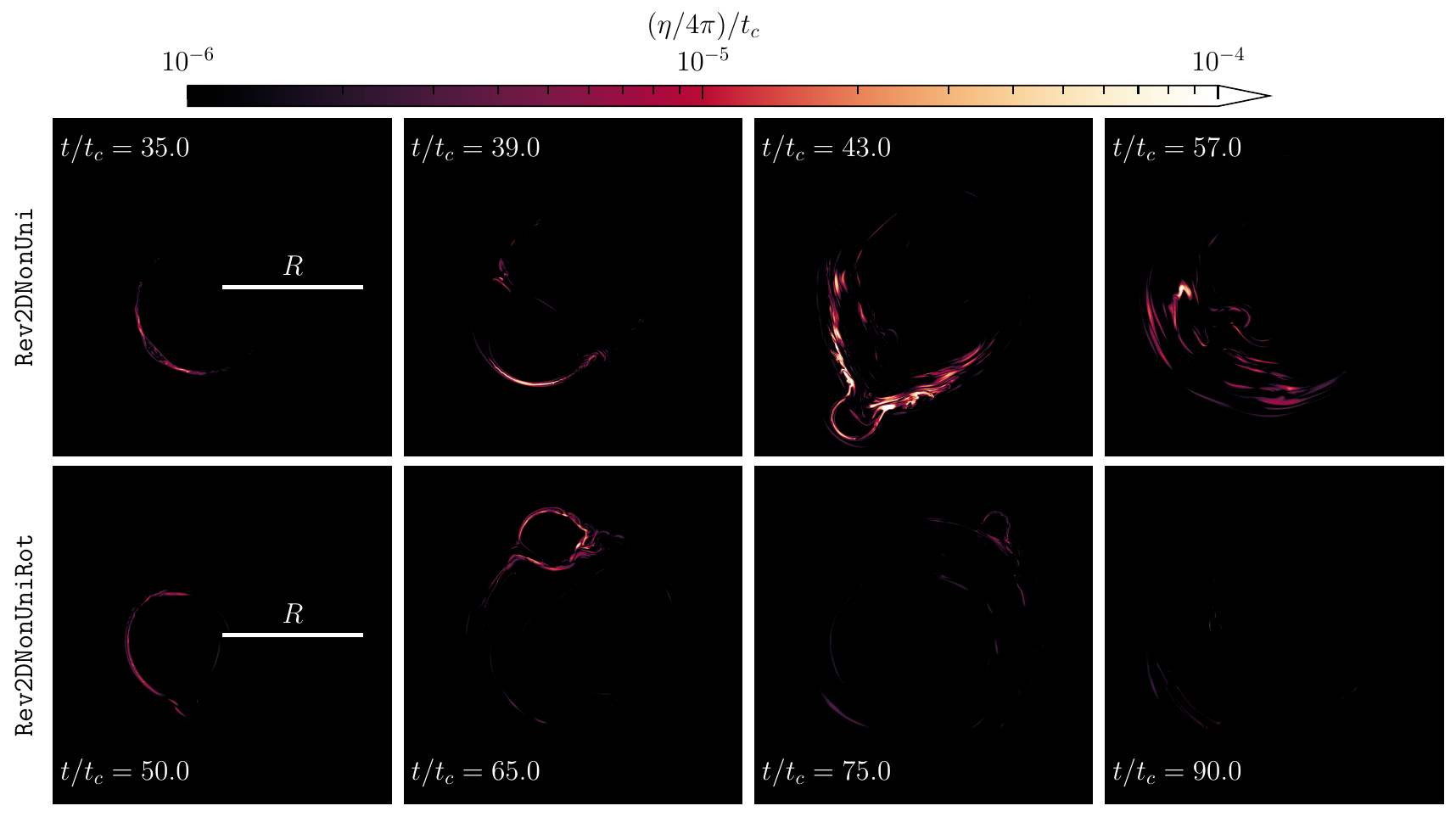} 
\caption{Time evolution of the non-uniform resistivity in simulations \texttt{Rev2DNonUni} and \texttt{Rev2DNonUniRot}. The resistivity remains localized to the dissipation regions and is largest in active sites of reconnection.} \label{fig:eta} 
\end{figure*}

\section{Non-Uniform Resistivity}\label{app:non-uni}

The non-uniform resistivity law that is adopted in simulations \texttt{Rev2DNonUni} and \texttt{Rev2DNonUniRot} is designed to capture the onset of charge starvation in regions where the available plasma cannot sustain the required current. It couples $\eta$ directly to the local state of the plasma (see Section~\ref{sec:non-uni}).  

The enhanced resistivity is mostly localized to regions undergoing reconnection. In particular, the thinning of the current sheet at an X-point drives the plasma toward charge starvation, reducing the available charge density (Figure~\ref{fig:rho}) and increasing the non-ideal electric field (Figure~\ref{fig:qEz}). Both effects act to increase $\eta$ (see Equation~\eqref{eq:eta}), providing enhanced dissipation precisely where the ideal-MHD approximation begins to break down. Conversely, in the surrounding upstream plasma, where the density remains close to its initial value and the non-ideal electric field is small, the resistivity remains close to its background, near-ideal value.

The spatial distribution of this enhanced resistivity is shown in Figure~\ref{fig:eta}. In both the (initially) static and rotating configurations, $\eta$ remains strongly localized within the actively reconnecting zone, rather than spreading throughout the flux tube. The upstream plasma is effectively ideal, and only a weaker enhancement in $\eta$ is seen within the broad, turbulent domain outside the main current sheet. These results indicate that this resistivity prescription applies to more dynamical situations than a planar current sheet \citep{Ripperda2026}.

\clearpage
\bibliography{ref}% Produces the bibliography via BibTeX.

\end{document}